\documentclass[a4paper,12pt]{article}

\usepackage[a4paper,left=1.2in,right=1.2in,top=1.15in,bottom=0.99in,nohead]{geometry}

\usepackage{setspace}

\usepackage{rotating} 
\usepackage{ntheorem}
\usepackage[normalem]{ulem}
\usepackage{csquotes}
\usepackage{amssymb}
\usepackage{lipsum}
\usepackage{amsmath}
\usepackage{latexsym}
\usepackage{dsfont}
\usepackage{multirow}
\usepackage{floatpag}
\usepackage{graphicx}
\usepackage{subfigure}
\usepackage{url}
\usepackage[authoryear]{natbib}
\usepackage{bbm}
\usepackage{braket}

\usepackage{authblk}
\usepackage{tikz}
\usepackage{relsize}
\usepackage{pdflscape}
\usepackage{longtable}
\usetikzlibrary{arrows}

\newcommand{\E}[1]{\mathbb{E}\left[#1\right]}

\newcommand{\tonde}[1]{\left(#1\right)}
\newcommand{\quadre}[1]{\left[#1\right]}

\newcommand{\accapo}{\nonumber\\}
\newcommand{\ra}{\rightarrow}

\newcommand{\capm}[2]{X_{#1,#2}^{ \mbox{\tiny{CAPM}}}}

\newcommand{\ba}{\begin{eqnarray}}
\newcommand{\ea}{\end{eqnarray}}
\newcommand{\bi}{\begin{itemize}}
\newcommand{\ei}{\end{itemize}}
\newcommand{\ben}{\begin{enumerate}}
\newcommand{\een}{\end{enumerate}}
\newcommand{\be}{\begin{equation}}
\newcommand{\ee}{\end{equation}}

\DeclareFontFamily{U}{mathx}{\hyphenchar\font45}
\DeclareFontShape{U}{mathx}{m}{n}{<-> mathx10}{}
\DeclareSymbolFont{mathx}{U}{mathx}{m}{n}
\DeclareMathAccent{\widebar}{0}{mathx}{"73}

\makeatletter
\newcounter{subhyp} 
\let\savedc@hyp\c@hyp

\newcommand{\normhyp}{%
  \let\c@hyp\savedc@hyp 
  \renewcommand\thehyp{\arabic{hyp}}%
} 
\allowdisplaybreaks

\begin{document}

\title{Assessing systemic risk due to fire sales spillover through maximum entropy network reconstruction\thanks{
This work is supported by the European Community H2020 Program under the scheme INFRAIA-1-2014-2015: Research Infrastructures, grant agreement $\#$654024 SoBigData: Social Mining and Big Data Ecosystem (http://www.sobigdata.eu).  We thank Fulvio Corsi and Piero Mazzarisi for inspiring discussions and useful suggestions.  Any mistake, substantial or formal, is our own responsibility.}}

\author[$\dagger$]{\normalsize{Domenico Di Gangi}}  
\author[$*$,$+$]{\normalsize{Fabrizio Lillo}}  
\author[$\dagger$]{\normalsize{Davide Pirino}}

\affil[$\dagger$]{Scuola Normale Superiore}
\affil[$*$]{Universit\'a di Bologna, Italy}
\affil[$+$]{Center for Analysis, Decisions, and Society, Human Technopole, Milano, Italy }

\maketitle
\begin{abstract}

Monitoring and assessing systemic risk in financial markets is of great importance but it often requires data that are unavailable or available at a very low frequency. For this reason, systemic risk assessment with partial information is potentially very useful for regulators and other stakeholders. In this paper we consider systemic risk due to fire sales spillovers and portfolio rebalancing by using the risk metrics defined by \cite{Greetal15}. By using a method based on the constrained minimization of the Cross Entropy, we show that it is possible to assess aggregated and single bank's systemicness and vulnerability, using only the information on the size of each bank and the capitalization of each investment asset. We also compare our approach with an alternative widespread application of the Maximum Entropy principle allowing to derive graph probability distributions and generating scenarios and we use it to propose a statistical test for a change in banks' vulnerability to systemic events.
\end{abstract}


\maketitle

\clearpage

\section{Introduction}

After the recent troubled years for the global economy, in which two severe crises (the $2007$ crisis of financial markets and the $2010$ sovereign debt crisis) have put the whole economic system in dramatic distress, vulnerability of banks to systemic events is now the main focus of a growing number of investigations of the academic community. Simultaneously, many research efforts are devoted to understand the role of banks or, broadly speaking, of financial institutions in the creation and in the spreading of systemic risk. Given the prominent importance of the topic and its multifaceted nature, the literature on evaluation and anticipation of systemic events is huge \citep[see][among many contributions]{KunDet98,KamCar99,Har09,Schetal09,Baretal10,DutCas11,Krietal11,Aletal12,Arnetal12,Bisetal12,Schetal12,Meretal13,Oetetal13}.

Several are the channels through which financial distress may propagate from one institution to another and, eventually, affect a vast portion of the global economy. Fire sales spillovers due to assets' illiquidity and common portfolio holdings are definitely one of the main drivers of systemic risk. Shared investments create a significant overlap of portfolios between couples of financial institutions. Such (indi- rect) financial interconnectedness is an important source of contagion, since partial liquidation of assets by a single market player is expected to affect all other market participants that share with it a large fraction of their own investments \citep[see][]{CorLilMar13,Huaetal13,Cacetal14,LilPir15}. Fire sales move prices due to the finite liquidity of assets and to market impact. In a perfectly liquid market there will be no fire sale contagion at all \citep[see][for a review on the role of liquidity in financial contagion]{AdrShi08}. Finally, leverage amplifies such feedbacks. In fact, as described in detail by \cite{AdrShi10,AdrShi14}, levered institutions continuously rebalance their positions inflating positive
and, most importantly, negative assets' price variations.

Assessing and monitoring systemic risk due to fire sales spillover is therefore of paramount importance for regulators, policy makers, and other participants to the financial markets. \cite{Greetal15} introduced recently a stylized model of fire sales, where illiquidity, target leverage, and portfolio overlap are the constituent bricks. They used the model to propose two systemic risk metrics:  systemicness and vulnerability of a bank. Given a market shock, the first is the total percentage loss induced 
on the system by the distress of the bank, whereas the second is the total percentage loss experienced by the bank when the whole system is in distress. In order to compute these quantities, a full knowledge of the portfolio composition of all banks is needed, because the systemicness and vulnerability of a bank depends on the portfolio and leverage of the other banks. 

\cite{Greetal15} applied their method to the European Banking Authority (EBA) data that resulted from the July 2011 European stress tests. These data provide detailed balance sheets for the $90$ largest banks in the European Union. \cite{DuaEis13} exploited a publicly available dataset of balance sheets of  US bank holding companies to apply the framework of \cite{Greetal15}.
They derive a measure of aggregate vulnerability that {\it [...] reaches a peak in the 
fall of 2008 but shows a notable increase starting in 2005, ahead of many other systemic risk indicators}.

In general, however, the detailed information set required to compute such systemic risk indicators might not be available. For example European stress test data are sporadic. Moreover the sampling frequency of balance sheet data is rarely higher than quarterly. Thus an important question is whether it is possible to estimate systemic risk due to fire sales spillovers in absence of data on portfolio composition of financial intermediaries. 

Two possible approaches have been proposed in the literature. The first one  \citep[see, among others, ][]{AdrBru11,Achetal12,BanDum15,Coretal15} is purely econometric and it is typically based on publicly available data on price of assets and market equity value of publicly quoted financial institutions. Generically the method consists in estimating
conditional variables, such as conditional Value-at-Risk or conditional Expected Shortfall. 
The econometric approach circumvents the 
unavailability of data on portfolio holdings, but pays this advantage with the introduction of a strong stationarity assumption: estimates 
based on the past information are assumed to be 
always good predictors of the future behavior of the system. Nevertheless, due to the 
nature of a global financial crisis, it is 
in the very moment of the onset of a period of distress that the stationarity 
assumption may fail to work properly. Moreover it is often restricted to publicly quoted institutions for which equity value are available at daily frequency. 

A second possible approach\footnote{There are, of course, many different approach 
to assess systemic risk in financial networks. For example, \cite{Amietal13} propose a 
rigorous asymptotic theory that allows to predict the spread of distress in interbank networks.}, followed in the present paper, 
consists in inferring the matrix of portfolio holdings using only a reduced, but easily available, information set, and/or deriving a probability distribution for the portfolio weights according to some criterion. This is typically achieved summoning the {\it maximum entropy principle} which postulates that \citep[][]{Anaetal13} {\it [...] subject to known constraints [...] the probability distribution that best represents our current knowledge and that is least biased is the one with maximal entropy}. The approach of Maximum Entropy, can be applied in at least two different ways that we distinguish clearly in the following, and is not new in systemic risk studies \citep[]{Mis11,Anaetal13,Musetl13,Squetal13,Baretal15}. It is widely used for inferring the structure of the interbank network when only data of total interbank lending and borrowing for each bank (plus possibly other information) are available \citep[for a comparison of different methods, see][]{anand2017missing, gandy2016bayesian}. 

The seminal contribution by   \cite{Mis11}, comparing the empirical Italian interbank network with that reconstructed via a Maximum Entropy optimization procedure, has shown that the latter is fully connected while the former is very sparse (see also \cite{Masetal12}) and,as a consequence of this misestimation, the reconstructed network underestimates the risk contagion\footnote{A complementary method is proposed by \cite{Anaetal15}.   Here the authors reconstruct the network of bilateral exposures for the German banking system via the matrix that, preserving some constraints, has the minimum density. Nevertheless, if cross entropy method underestimates systemic risk by overestimating the network density,  \cite{Anaetal15}  show that, for a similar reason, minimum density returns positively biased estimates. Hence, the two approaches can be used jointly together to create a corridor in which the true systemic risk should lay.}. Recently a comparison of network reconstructions techniques has been carried out also for bipartite networks \citep[][]{ramadiah2017reconstructing}.

A part from network reconstruction, the use of entropic methods is widespread in economic sciences. For example, it is widely used 
in econometrics for the estimation 
of probability densities,  as it is witnessed by a vast stream of contributions in this direction \citep[see, among others, the contributions by][]{ZelHih88,Ryu93,Wu03,KouPie04,ParBer09,Ilhetal11,Che15}. An interesting point of coomparison for our paper is the use of entropy in the theory of portfolio choice\footnote{We thank an anonymous referee for this suggestion.}. When investors are uncertain about the probability structure of reality and, being averse to ambiguity, use the relative entropy  as a way of penalizing \citep[see, among the most notable contributions
in this field, the works by][]{HanSar01,Macetal06,BrePar08,Zhoetal08,gilmar11,Zhoetal13}. 
In this respect it is important to clarify that we implicitly adopt the maximum entropy principle from the point of 
view of a regulator (or a social planner) who, irrespectively of the decisional process 
which is behind the creation of the network, is solely interested in having 
an unbiased estimate of systemic risk. Hence, our perspective could be thought as 
orthogonal to that adopted by the ambiguity aversion literature, in the sense that
our declination of the maximum entropy principle is purely inferential and it is not meant to mimic, in any way, 
the banks' decisional processes that have created the network and, accordingly, the prevailing level of
 systemic risk. 

In this paper we propose to apply maximum entropy approach to the inference of the network of portfolio weights in order to estimate metrics of systemic risk due to fire sales spillovers. Specifically, we show how indirect vulnerability, systemicness  \citep[as defined by][]{Greetal15} and the aggregate systemic risk of US commercial banks can be estimated when only a partial information (the size of each bank and the capitalization of each asset) is available. Differently from the interbank studies \citep[as in][]{Mis11,Masetal12,Anaetal15}
we deal with bipartite networks, namely graphs\footnote{Throughout all the manuscript we use the terms \enquote{network} and \enquote{graph} interchangeably.} 
whose nodes can be divided into two sharply distinguished sets that,
in our case, are commercial banks and asset classes. More specifically, we analyze the quarterly networks of US commercial banks'
exposures in the period 2001-2013 using the {\it Federal Financial Institutions Examination Council} (FFIEC) through the {\it Call Report} files\footnote{Hence our dataset 
is quite similar to that adopted by \cite{DuaEis13}, but it profits from a larger sample of banks, since commercial banks
are fairly more numerous than bank holding companies.  All data are available at: \url{https://cdr.ffiec.gov/public/} }. 
We compute, for each quarter, systemicness and vulnerability of each bank and the aggregate vulnerability of the system. We compare them with the values inferred assuming the balance sheet compositions of the banks were not known. In this sense our paper is similar to \cite{Mis11}, but applied to systemic risk due to fire sale spillover rather than to cascades in the interbank network. Differently from the interbank case, we find that newly introduced maximum entropy methods are very accurate in assessing systemic risk due to fire sales spillover when partial information is available.

The contribution of this paper is divided into two main parts. First, following a practice that is largely diffused among researchers of both academic institutions and central banks 
\citep[see, among others,][]{SheMau98,UppWor04,Wel04,Mis11,Sac14},  we reconstruct the matrix of portfolio holdings as such that minimizes the cross entropy (or Kullback-Leibler divergence) from a initial guess. Despite this approach has often been referred to as maximum entropy, or matrix balancing, in order to avoid confusions with different methods discussed in the following, we refer to it as Cross-Entropy method. We show that this approach does a very good job in our case, providing unbiased estimates of the systemic risk metrics defined by  \cite{Greetal15}.  Besides, we show that the reconstructed matrix corresponds to that implied by the Capital Asset Pricing Model, hence it possesses a clear economic meaning.

Second, we compare Cross-Entropy with a different approach to entropy maximization, which allows to define a probability mass function for graphs (ensemble) by maximizing entropy under suitable constraints where some average quantities are set equal to the ones observed in data. Despite the economic intuition of this approach is less sharp than the previous one, the method is widespread in the literature and allows performing scenario generation. We propose a new ensemble, termed MECAPM, which (i) satisfies a set of economically motivated constraints, (ii) behaves in average as the cross entropy method proposed before, and (iii) allows for scenario generation, potentially useful for supervisory authorities to test if a specific institution has increased its systemicness with respect to the past.

We structure our paper as follows. Section \ref{sec:vulsys} introduces some nomenclature and briefly describes the risk metrics
of \cite{Greetal15}. The dataset of US commercial banks provided by the FFIEC is discussed in Section \ref{sec:data}. In Section \ref{sec:cross_entropy} we present the cross entropy method and show its performances for the estimation of systemic risk. In Section  \ref{sec:mod_recns} we compare the cross entropy method with the maximum entropy alternative which derives a probability distribution of graphs. This is useful, among other things, to introduce a statistical test for surveillance activities by central banks and other regulatory institutions. Finally Section \ref{sec:conc} summarizes the main contributions of the paper. Appendices provide additional information 
on the construction of the dataset of bank portfolio holdings and  all the analytical computations omitted in the main text. 

\section{Systemic risk metrics: Vulnerability and Systemicness}\label{sec:vulsys}

In this paper we use some metrics of systemic risk due to fire sales, which have been recently introduced by 
\cite{Greetal15}. They consider a system composed by $N$ banks and $K$ asset classes. Portfolio holdings are described by the  $N\times K$ matrix $\mathbf{X}$, 
whose element $X_{n,k}$ is the dollar-amount of $k$-type assets detained by bank $n$. The corresponding 
matrix of portfolio weights is thus 

$$
W_{n,k} \tonde{\mathbf{X}}= \frac{X_{n,k}}{\sum_{k^{\prime}=1}^{K}X_{n,k^{\prime}}}.
$$ 

In what follows, we introduce a discretization of the elements of $X$'s, 
in such a way that the matrix $\mathbf{X}$ belongs to the space $\mathbb{N}^{N\times K}$
of $N\times K$ integer valued matrices. 
In the empirical application we will use the resolution of the dataset which is $10^3\$$.

The total asset size $A_{n}$ of the $n$-th bank and the total 
capitalization\footnote{More precisely, the quantity $C_{k}$ is the total amount of asset's $k$ capitalization due to the banking sector. To simplify the notation we will call it capitalization.} $C_{k}$ of the $k$-th asset class are easily computed as, respectively,
the total row and column sums of the matrix $X_{n,k}$, in formula
\be\label{eq:rowcolsum}
A_{n}\tonde{\mathbf{X}} = \sum_{k=1}^{K}X_{n,k},~~~~~~C_{k}\tonde{\mathbf{X}} = \sum_{n=1}^{N}X_{n,k},
\ee
where we have explicitly expressed the dependence of $A_{n}$ and $C_{k}$ from $\mathbf{X}$. 

The rectangular matrix $\mathbf{X}$ can be naturally associated to a bipartite network, i.e. a graph whose vertices 
can be divided into two disjoint sets such that every 
edge connects a vertex in one set to 
one in the other set, the two sets being the banks and the asset classes. 
In the network jargon, $A_n\tonde{\mathbf{X}}$ and $C_k\tonde{\mathbf{X}}$ are called the strength sequences.

A relevant information concerning 
the balance sheet of each bank 
$n$ is the total equity $E_n$, from which one can compute the leverage as $B_n=\frac{A_n-E_n}{E_n}$ \citep[as in][]{Greetal15}. 
Finally, each asset class is characterized by an 
illiquidity parameter $\ell_k$, with $k=1,...,K$, defined as the return 
per dollar of net purchase 
of asset $k$\footnote{The assumption 
of linear price impact comes directly from the framework of \cite{Greetal15}. 
Although a square-root law
fits the data better, the linear assumption 
has been widely adopted in the literature 
 \citep[see][among others]{Gatetal12,ConWag14,LilPir15} 
 and has been empirically validated at daily frequency by \cite{Obz08}.}.
 
This setting is used in \cite{Greetal15} to 
define three metrics of systemic risk, 
capturing the effect of fire sales in response to a shock on the price of the assets. 
This is described by the $K$ dimensional vector $-\boldsymbol{\varepsilon}=\tonde{-\varepsilon_1,...,-\varepsilon_{K}}$, 
whose components are the assets' shocks. They define:
\begin{itemize}
\item  {\bf Aggregate vulnerability} $AV$ as {\it [...] the percentage of 
aggregate bank equity that would be wiped out by bank deleveraging if there 
was a shock [...] to asset returns}. 
\item {\bf Bank systemicness} $S_n$ as {\it  the contribution of bank $n$ to aggregate vulnerability}.
\item {\bf Bank's indirect vulnerability} $IV_n$ as {\it [...] the impact of the shock on its equity through the deleveraging of other banks}.
\end{itemize}  

By assuming that banks follow the practice of leverage 
targeting and that, in response to a negative asset shock, 
they sell assets proportionally to their pre-shock portfolio holdings, \cite{Greetal15}  
show that $S_n$ can be decomposed as
\be\label{eq:Sn}
S_n = \Gamma_n\,\frac{A_n}{E}\,B_n\,r_n,
\ee
where $E$ is the total equity, $E=\sum_{n=1}^{N}E_n$, $r_n$ is the $n$-th element of the vector $\mathbf{r}=\mathbf{W}\,\boldsymbol{\varepsilon}$, i.e. the portfolio return of bank $n$ due to the shock $\boldsymbol{\varepsilon}$, and 
$$
\Gamma_n = \sum_{k=1}^{K}\tonde{\sum_{m=1}^NA_m\,W_{m,k}}\,\ell_k\,W_{n,k}.
$$
The aggregate vulnerability is computed simply as 

\be\label{eq:AV}
\textrm{AV}=\sum_{n=1}^{N}S_n.
\ee 

Finally, the  indirect vulnerability of a bank is
\be\label{eq:IVgreen}
\textrm{IV}_n  = \tonde{1+B_n}\,\sum_{k=1}^{K}\ell_k\,W_{n,k}\,\sum_{n^{\prime}=1}^{N}W_{n^{\prime},k}\,A_{n^{\prime}}\,B_{n^{\prime}}\,r _{n^{\prime}}.
\ee

In what follows we often assume, as in \cite{DuaEis13},  that $\epsilon_k=1\%$ for all $k=1,...,K$, which in turns implies that 
$r_n=1\%$ in equations \eqref{eq:Sn} and \eqref{eq:IVgreen}. Note however that if all the assets are shocked by the same amount, our results do not depend on it, since the systemic 
risk measures will have only a different pre-factor. In Section \ref{sec:shocks} we consider other shock scenarios to test the robustness of our methods. Finally, 
we set the liquidity parameter at  $\ell_k=10^{-10}$ for all asset classes 
except for cash, for which we put $\ell_k=0$ \citep[as in][]{Greetal15,DuaEis13}. As a final comment it should be noted that  \citep{Greetal15} add two more constraints to the problem. First, when direct losses of a bank exceed its equity, the bank liquidates all the assets. Second, leverage is capped to the value $30$. In our empirical investigation we have followed \citep{DuaEis13} who do not add these constraints. We have however compared the aggregate vulnerability of the US banking system (see next Section for the data used) under the two model specifications. We have found that the difference is less than 1\% with the exception of few quarters around the end of 2009 when it reaches 10\%.

 It is important to stress that the \cite{Greetal15} method to estimate systemic risk metrics is essentially static. As it is standard in stress testing, a given scenario of price changes at a given time is considered and then, given the balance sheet and portfolio composition of banks at that time, the consequences of deleveraging and fire sales are computed. Thus no past information (even when available) on balance sheets or prices is ever used in the methodology. This is of course a limitation since the decision on how to deleverage in a certain quarter depends in reality also on past market price behavior as well as on deleveraging in the last quarters. Such an extension, although interesting, is beyond the scope of \cite{Greetal15} model as well as of the vast majority of stress test methods. It would require to choose scenarios containing the price changes at more than one quarter as well as being able to disentangle the price changes due to fundamental reasons from those due to past deleveraging.  The definition of a dynamic stress test is clearly beyond the scope of our paper and we will stick to the standard static stress test approach. As in \cite{DuaEis13}, in the empirical application below we will consider a stress test for each available quarter, discarding all the information coming from past quarters. Thus even if apparently we are treating a time series of portfolios of length $T$, as a matter of fact we are repeating $T$ times the (static) stress test.

In the next section we present the dataset that we use in our analysis to 
measure systemic risk, as captured by the metrics of \cite{Greetal15}, in the US banking sector. 
Such a dataset allows us to have quarterly estimates of systemicness, aggregate, and indirect vulnerability 
and to compare these estimates with those inferred from the Cross-Entropy approach and the Maximum Entropy principle.  
Since we have to deal with both real and reconstructed (or sampled form a statistical ensemble) networks, from now on we follow the convention to add a superscript $x^{\star}$ to any variable $x$ whenever it is referred to a real (observed) network, while the variable $x$ is represented 
without the superscript $\star$ every time it is referred to a reconstructed network (e.g. one 
sampled from a statistical ensemble as described in Section \ref{sec:mod_recns}).

\section{Data}\label{sec:data}

All regulated financial institutions in the United States are required to file periodic financial information with their incumbent regulators. 
The {\it Federal Financial Institutions Examination Council},  is the regulatory institution responsible to collect and maintain the data used in our analysis. 
The financial institutions subject of our investigation are {\it Commercial Banks and Savings and Loans Associations}.
The FFIEC defines officially a commercial bank as\footnote{See \url{http://www.ffiec.gov/nicSearch/FAQ/Glossary.html}.}: \enquote{\textit{[...] a financial institution that is owned by stockholders, operates for a profit, and engages in various lending activities}}. FFIEC requires commercial banks to 
file the quarterly {\it Consolidated Report of Condition and Income}, generally referred to 
as {\it Call Report}. Each bank is required to fill a form with detailed 
information on its financial status, in particular on its balance sheet. 
The specific reporting requirements depend 
upon the size of the bank and whether or not it has 
any foreign office. 
The form FFIEC031 is used for banks with both domestic (U.S.) and foreign (non-U.S.) 
offices while form FFIEC041 is designed for banks with domestic (U.S.) offices only.
A {\it Saving and Loan Association} is a financial institution 
that accepts deposits primarily from individuals and channels its funds primarily into residential mortgage loans.
From the first quarter of 2012, all {\it Savings and Loan Associations} are required to file the same reports, thus they are included in the dataset since then.

\begin{figure}[t]
  \begin{center}
\includegraphics[width=7cm,height=5cm]{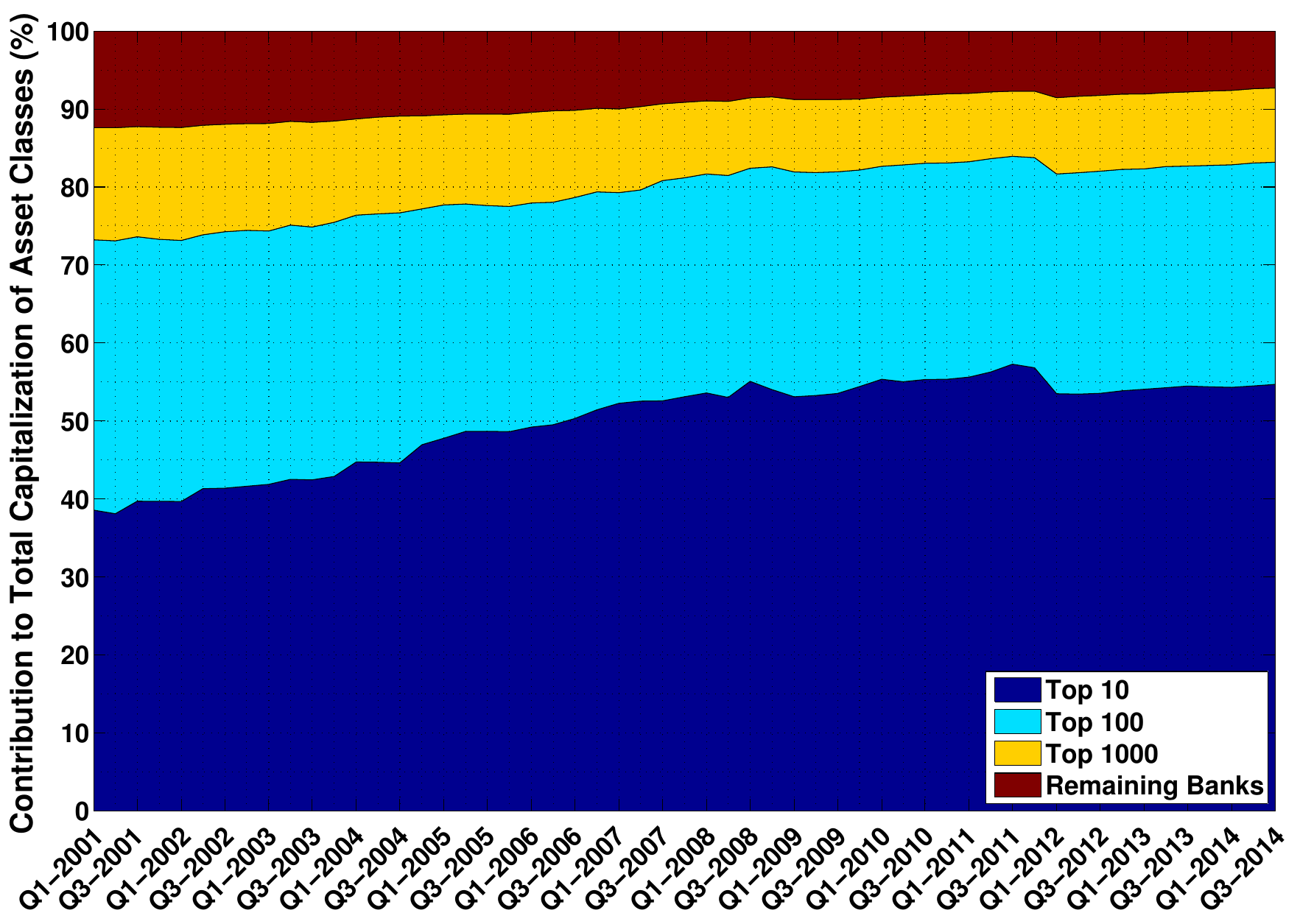}
\includegraphics[width=7cm,height=5cm]{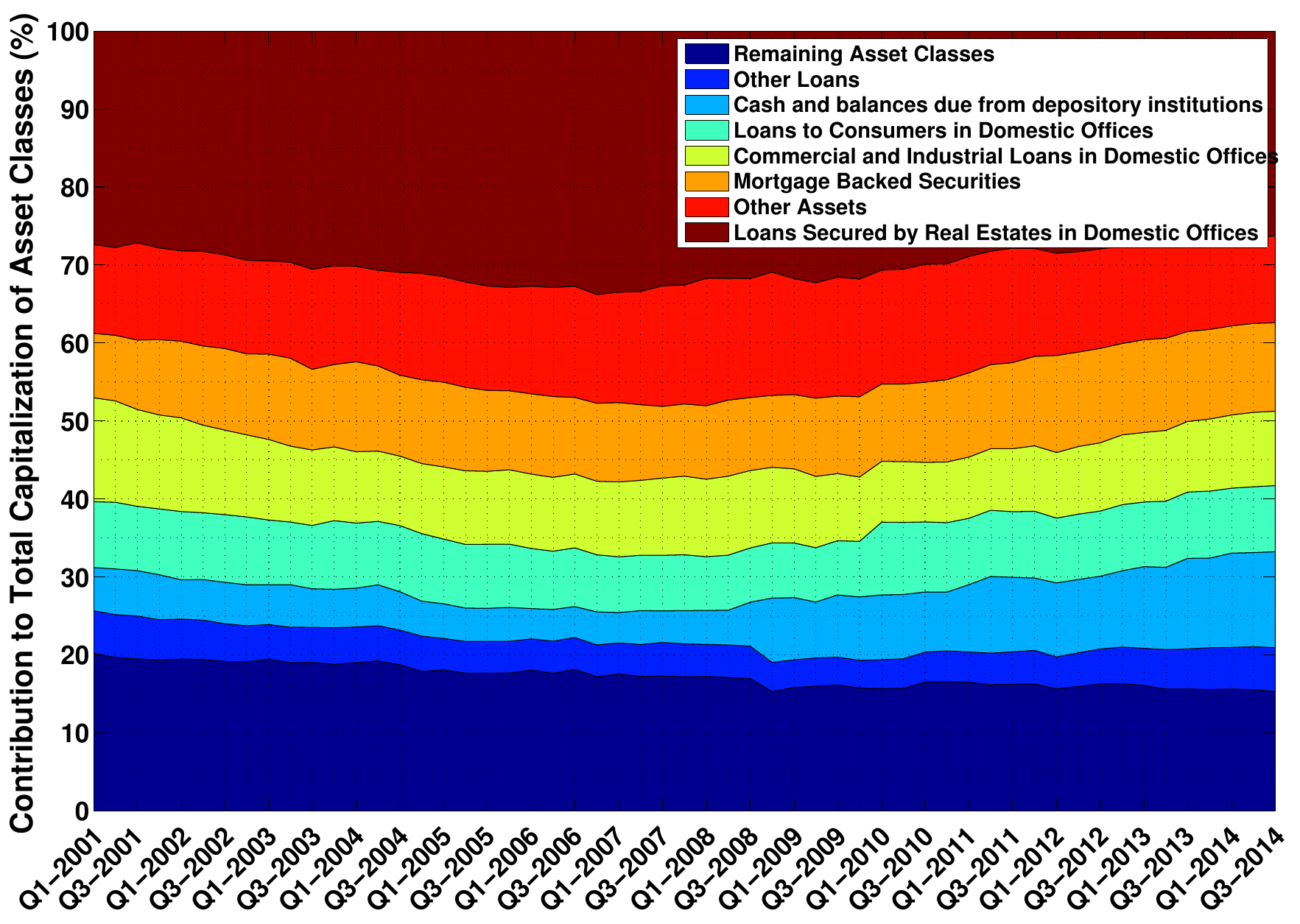}
\caption{\small{We report in the left panel the percentage of total assets detained by top 10, top 100, top 1000 and the remaining banks in shaded areas of different colors. 
A vast portion of total assets is controlled by the top 10 banks. In the right panel we report, for each quarters, the contribution of the top seven asset classes (in terms of capitalization) 
to the total capitalization. A large percentage to total asset capitalization is due to Loan Secured by Real Estates in Domestic Offices.}}\label{fig:data_description}
\end{center}
\end{figure}

The data provided by the {\it Call reports} are publicly available\footnote{\footnotesize{See \url{https://www.chicagofed.org/banking/financial-institution-reports/commercial-bank-data}.}}
 since 1986, although the form changed considerably throughout  the years, showing an increasing level of details requested. 
To have a good compromise between the fine structure of data and a reasonably populated statistics  we considered the time period going from March 2001 to September 2013,
for a total of $55$ quarters. The number of financial institutions present in the data 
is pretty stable during quarters, starting from  approximately $9,000$ entities in the first quarter 
and ending in roughly $6,500$ in the last one. 
The asset categories have been created as coherent sums of codes.
We describe the procedure adopted to form asset classes in Appendix \ref{app:appendix_data_description} along with some data statistics.  
In particular, we aggregate data in a set of 20 asset classes following the rationale of \cite{DuaEis13}, that is each of the 20
asset classes is composed in such a way that, in case of a fire 
sale of assets belonging to a specific class, the 
price impact would be restricted mainly to 
the assets in the same class. In other words, it is reasonable to assume that the co-illiquidity (or cross impact)
of two different asset classes is negligible. 
The twenty macro asset classes used to build the network are 
described in Table \ref{tab:ass_com} of Appendix \ref{app:appendix_data_description}, which also 
documents in detail how they have been formed.
In the left panel of Figure \ref{fig:data_description} 
we show how the total asset value is concentrated on the top tiered banks. 
 The right panel of Figure \ref{fig:data_description} shows the relative 
importance of the top seven assets classes (in terms of total capitalization), revealing that a large portion 
of the total capitalization is due to {\it Loan secured by real estates in domestic offices}.  

\begin{figure}[t]
  \begin{center}
\includegraphics[width=0.8\textwidth]{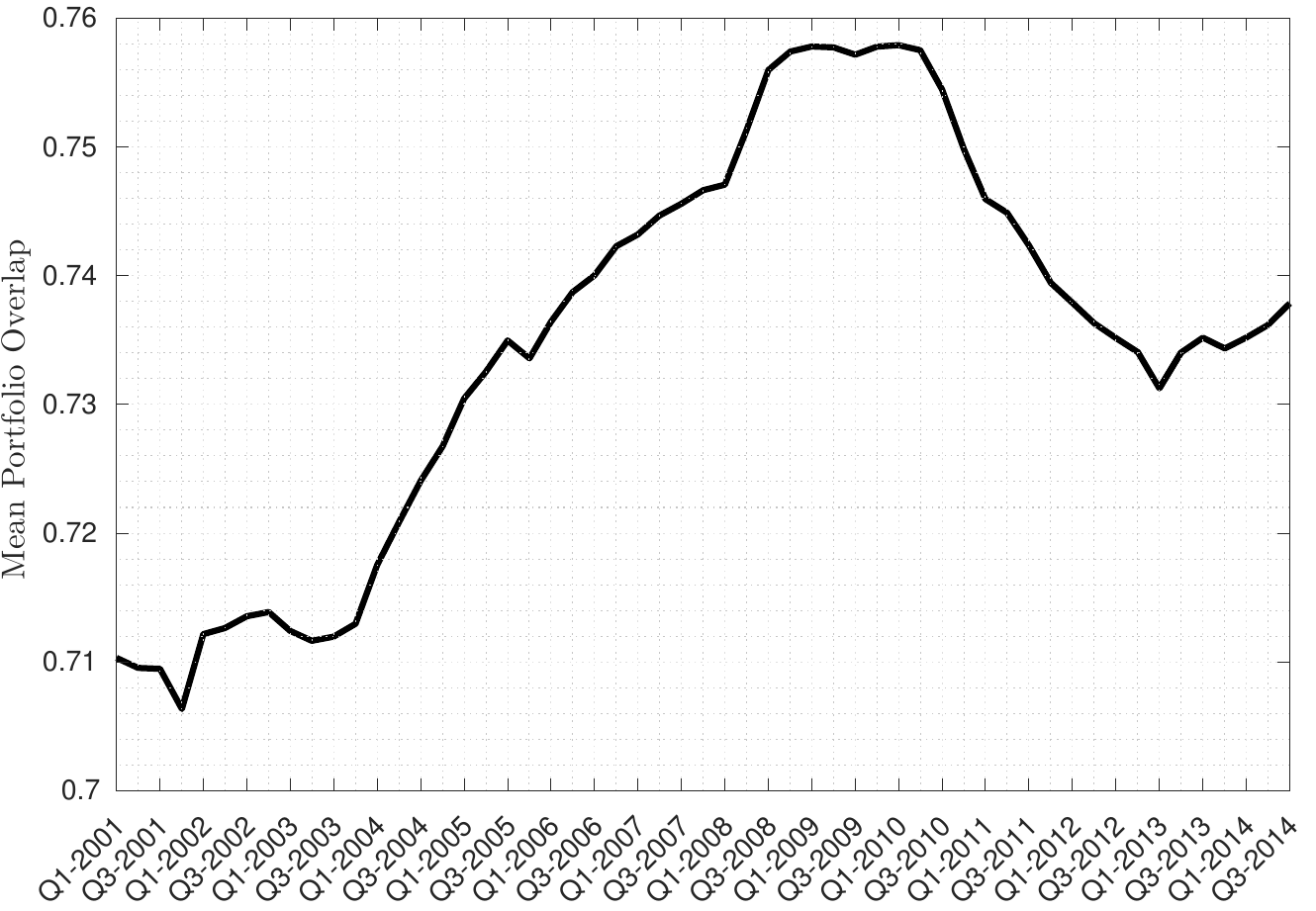}
\caption{\small{Mean cosine (or $L_2$) similarity of pairs of portfolios of US commercial banks.}}\label{fig:overlap}
\end{center}
\end{figure}

To test the role of portfolio similarity in systemic risk, we report in Figure \ref{fig:overlap}  the time series of the mean similarity between all the pairs of banks' portfolios. Similarity is measured with the cosine (or $L_2$) norm i.e. the cosine of the angle formed by the two vectors $\{X^*_{n,k}\}_{k=1,..,K}$ and $\{X^*_{m,k}\}_{k=1,..,K}$ representing the portfolios of bank $n$ and $m$, respectively. 
The plot shows clearly that similarity between portfolios has increased significantly before the 2008 crisis, making the systemic risk higher. A pretty similar pattern is shown by the aggregate vulnerability in Figure \ref{fig:agg_sys}, we will turn later on this point.

In conclusion, for each quarter we are able to construct 
a matrix $\mathbf{X}^{\star}$ of bank holding whose element $X_{n,k}^{\star}$ is the 
total dollars invested by the $n$-th bank in the $k$-th asset class. It is important to note that the matrix $\mathbf{X}^{\star}$ has around $50\%$ of   zero entries. Thus the network is relatively dense, but far from being fully connected. Simply put, 
the portfolio of the typical bank in the dataset does not contain investments in all the $20$ assets classes.

\section{The Cross-Entropy Approach for Systemic Risk Assessment}\label{sec:cross_entropy}
 
Cross-Entropy is a  method\footnote{The method is also known as matrix balancing, or maximum entropy matrix reconstruction. We refer to it as Cross-Entropy in order to clearly distinguish it from a method deriving probability distributions over all possible matrices (graphs). The latter is introduced in Section \ref{sec:mod_recns} as a competing approach for systemic risk reconstruction, and we refer to it as Canonical Maximum Entropy Ensemble method.}, largely adopted by scholars and researchers of central banks, used to reconstruct 
a target matrix  (as the inter-bank matrix) from partial knowledge of its properties. The idea is to select an {\it a priori} guess for the matrix and then to find its closest matrix subject to some constraints. 
In the simplest case, such constraints are non negativity conditions of matrix elements and the total row and column sums. Finally, as a measure of distance to be minimized between the guess and the target matrix one uses the Kullback-Leibler divergence (also called relative entropy).  

For the specific case of the system of bank holdings for US commercial banks we assume to 
have at our disposal, for each quarter, only the information on the total asset size $A_n^{\star}$
for the $n$-th bank and the total capitalization $C_k^{\star}$ for the $k$-th asset class. The Cross-Entropy approach derives the target matrix
$\mathbf{X}$ as that which solves the optimization  problem

\begin{equation}\label{eq:kubl}
\begin{aligned}
& \underset{\mathbf{X}}{\text{min}}
& & \sum_{n=1}^{N}\sum_{k=1}^{K}  X_{n,k}\,\log\tonde{\frac{X_{n,k}}{\widetilde{X}_{n,k}}} \\
& \text{s.t.}
& &  \sum_{n=1}^{N} X_{n,k} = A_n^{\star},~n=1,...,N,\\
& & &\sum_{k=1}^{K} X_{n,k} = C_k^{\star},~k=1,...,K,\\
& & & X_{n,k}\geq 0,
\end{aligned}
\end{equation}
where $\widetilde{X}_{n,k}$ are the entries of a given guess 
matrix. Note that the cases analyzed in the interbank lending literature  \citep{Mis11}
typically have an additional constraint that diagonal elements vanish, required
to avoid a single institution to be simultaneously a borrower and lender to itself \citep[see, for example, the Appendix B in][]{Mis11}. 
The matrix of portfolio holdings analyzed here does not require any of such kind of restrictions. 

We suggest to use the capital asset pricing model (CAPM) to form an economically motivated initial guess. In a standard CAPM, 
investors choose their portfolio in such a way that 
each weight on a stock is the fraction of that stock's market 
value relative to the total market value of all stocks \citep[][]{Sharpe64,Linter65,Mossin66}.  
Since $A_n^{\star}$ is the total asset size 
of the $n$-th bank and since the total market value of all stocks is given by $L^{\star}=\sum_{k=1}^{K}C_{k}^{\star}$, the  portfolio weights expected by CAPM are given by\footnote{More precisely, we are deriving what can be addressed as a \enquote{banking-CAPM} since, as mentioned before,  $C_k^{\star}$ represents the total amount of asset's $k$ capitalization due to the banking sector and not its total market capitalization.}
\be\label{eq:capm}
\capm{n}{k} =\frac{C_k^{\star}}{L^{*}}\,A_n^{\star}. 
\ee
Notice that this choice of the initial guess is the same used in \cite{Mis11} for the interbank market, even if in that case the CAPM interpretation is less direct. Given that in \eqref{eq:kubl} the condition on the diagonal 
elements is absent and since the  Kullback-Leibler divergence is always positive, 
the optimal solution of the Cross-Entropy problem in \eqref{eq:kubl} when $\widetilde{X}_{n,k}=\capm{n}{k}$
is nothing but the $\capm{n}{k}$ itself. To distinguish the estimator from the Capital Asset Pricing Model of \citep[][]{Sharpe64} we will call the former Cross Entropy CAPM (CECAPM) estimator. Note that thanks to the bipartite nature of
the network under study we do not have to resort
to numerical routines to solve problem \eqref{eq:kubl}.  If other constraints are added to the problem (e.g. that some banks cannot invest in some asset classes) one could numerically solve the problem \eqref{eq:kubl} with the additional constraints.\\

\subsection{Assessing aggregate vulnerability}\label{sec:aggregate}

\begin{figure}[!ht]
\begin{center}
\includegraphics[width=0.8\textwidth]{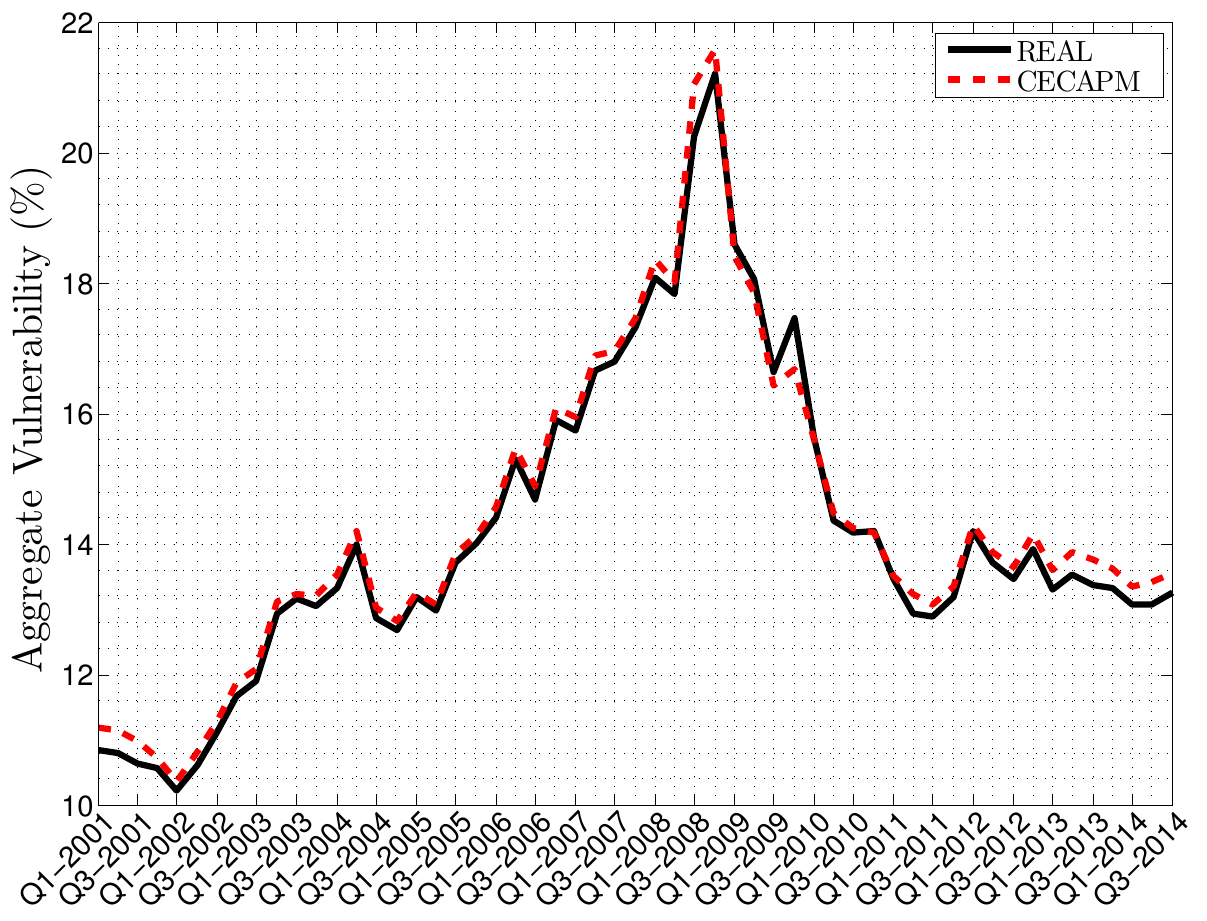}
\caption{This figure reports as a black continuous line the aggregated vulnerability, as 
defined by equation \eqref{eq:AV}, computed on the matrix $X_{n,k}^{\star}$ of portfolio 
holdings as provided by the FFIEC dataset of US commercial bank holdings, described in Section \ref{sec:data}. 
The dashed red line refers to the aggregate vulnerability reconstructed with CECAPM.}\label{fig:agg_sys}
\end{center}
\end{figure}

We now empirically test the validity of the cross entropy approach in estimating the aggregate vulnerability on our data.

Figure \ref{fig:agg_sys} compares the true value of the aggregated vulnerability, obtained by using the real matrix of portfolio compositions, with the one obtained with the cross entropy method. It is clear that CECAPM provides estimates of AVs in excellent agreement with the real one, in spite of the fact that the true portfolio matrix is quite different from that of CECAPM, because in the former roughly half of the matrix elements are zero while the latter models have adjacency matrices with all non vanishing elements.

An important implication of Figure \ref{fig:agg_sys} 
is that, at least for the dataset under analysis, it is not necessary to 
know the matrix $\mathbf{X}^{\star}$ to assess the systemic risk as measured by the aggregate vulnerability. The knowledge of banks' size and assets' capitalization is enough to infer the matrix $\capm{n}{k}$, which very
well reproduces the aggregate behavior (in terms of systemicness) of the system. 
This is different from the result of \cite{Mis11} for the interbank network, since he finds that the Cross-Entropy 
approach significantly underestimates systemic risk, while in our case
the bias is negligible.

\subsubsection{Robustness to different shock scenarios}\label{sec:shocks}

\begin{figure}[t]
\begin{center}

\includegraphics[width = 0.49\textwidth]{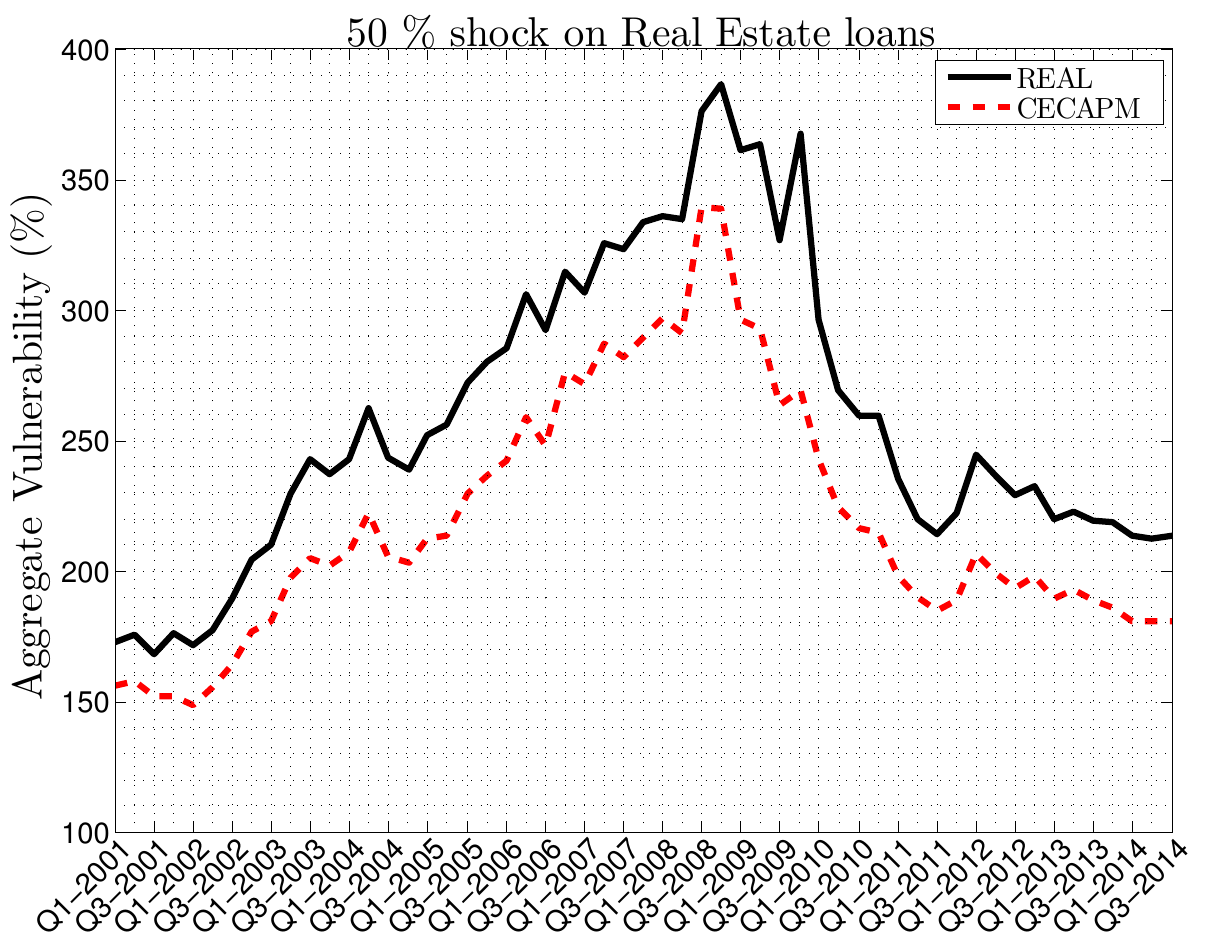} 
\includegraphics[width =  0.49\textwidth]{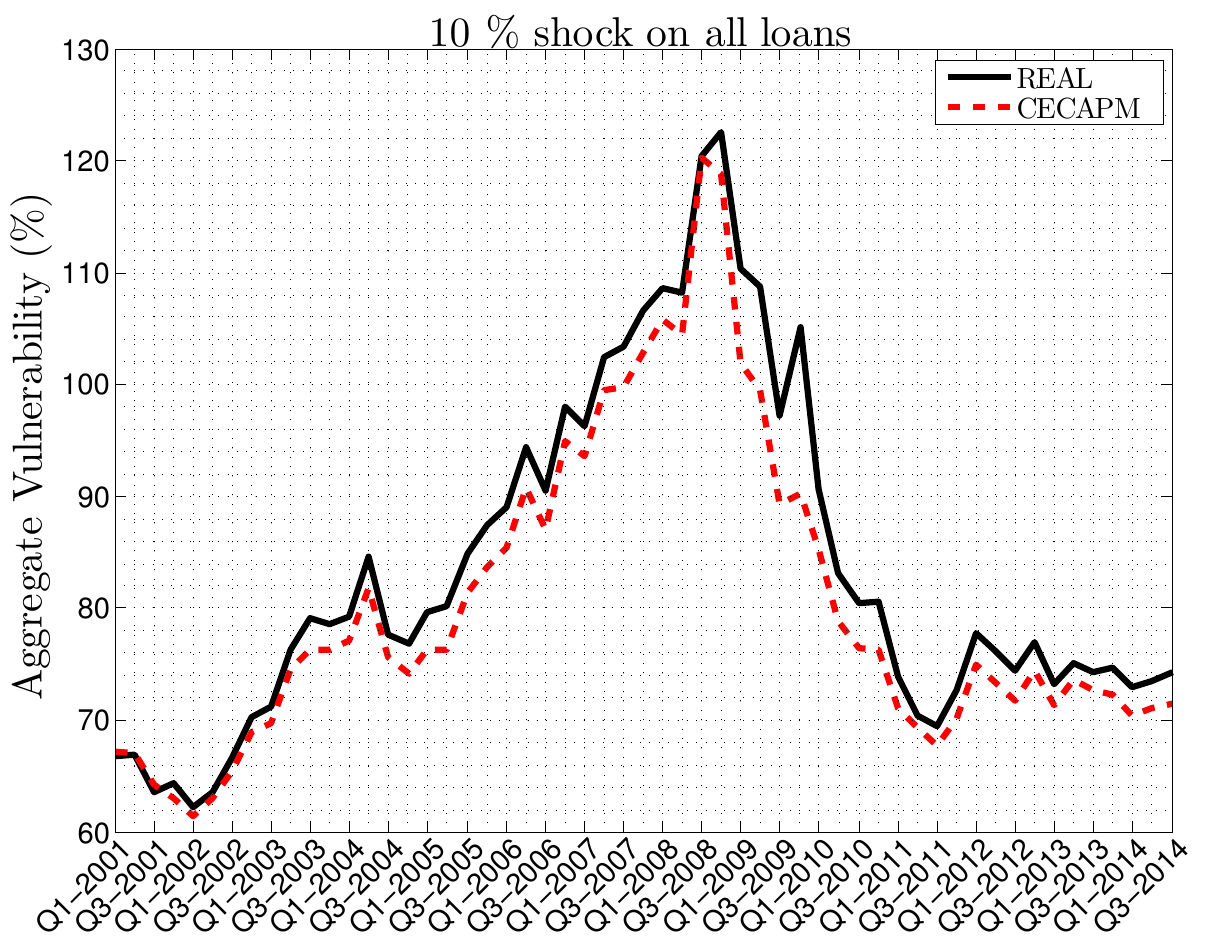} \\
\includegraphics[width = 0.49\textwidth]{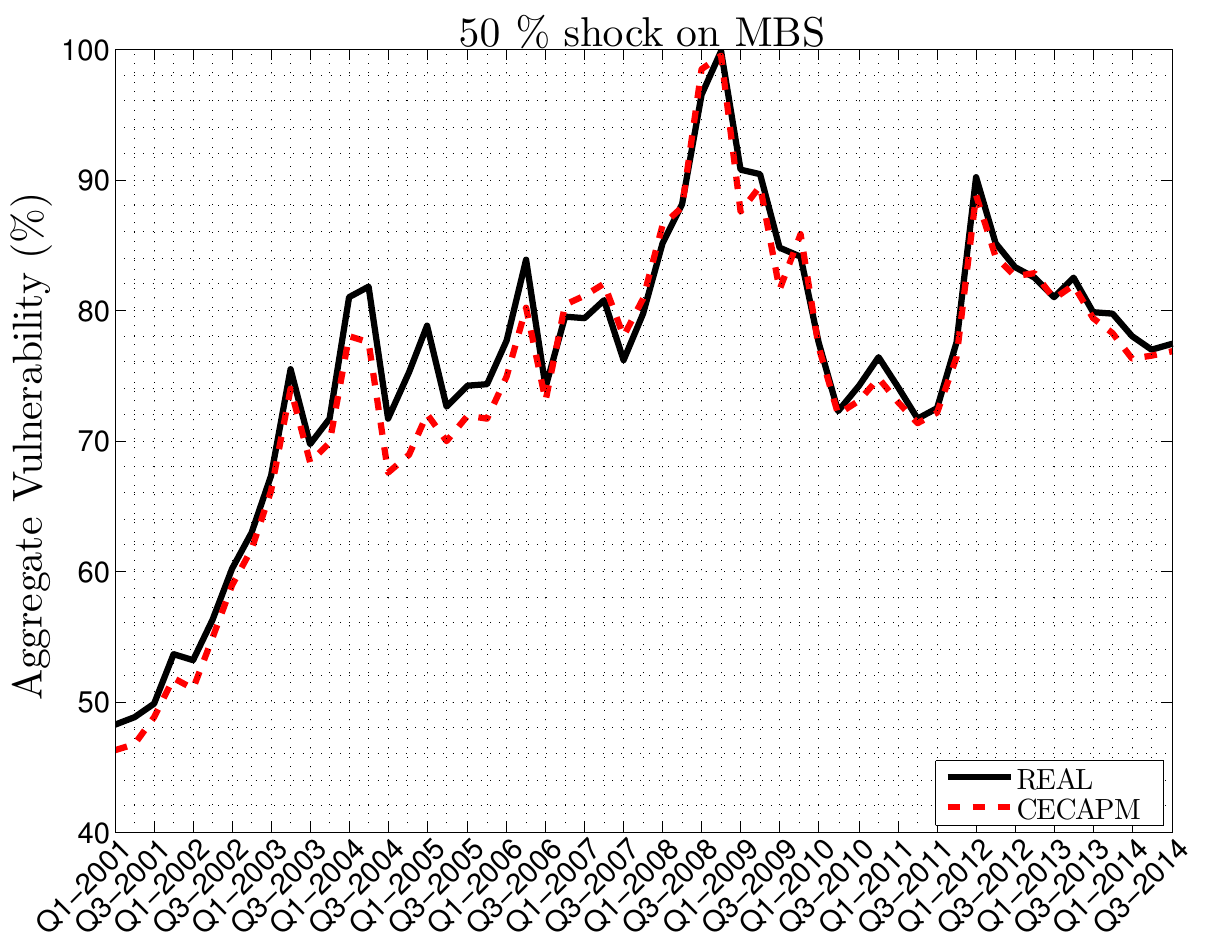} 
\includegraphics[width =0.49 \textwidth]{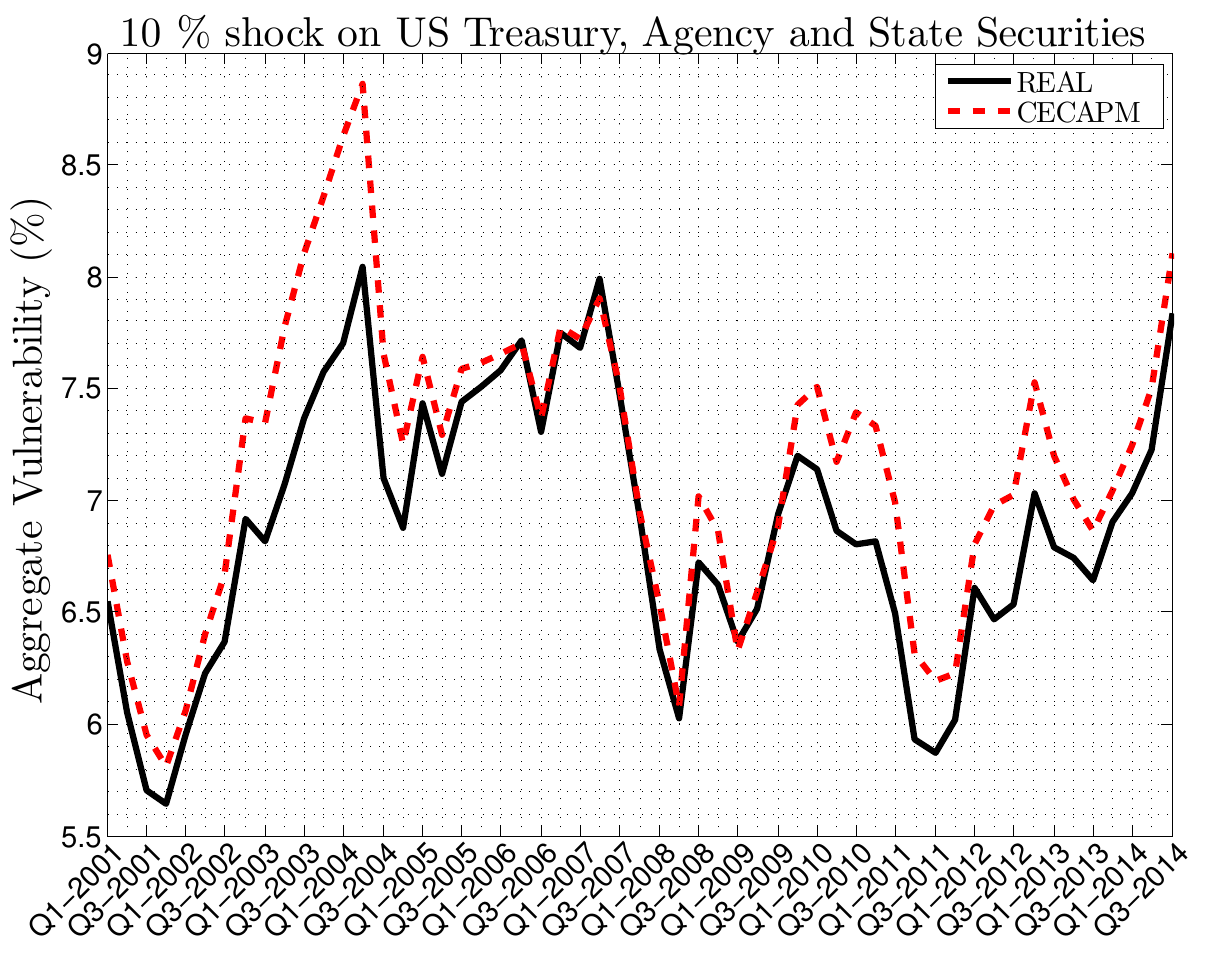} \\
 \caption{Aggregate vulnerability under different shock scenarios. Each panel reports the AV obtained from the full knowledge of portfolios composition and those obtained using the CECAPM recontruction.}\label{fig:othershocks}
\end{center}
\end{figure}

The estimation and reconstruction of AV has been performed by assuming a uniform shock of 1\% across all asset classes. However our results are robust also to other shock scenarios. To show this, we have repeated the above analysis by considering other cases, namely: (i) a 50 \% shock on Real Estate loans (2 asset classes); (ii) a 10 \% shock on all loans (8 asset classes); (iii) a  50\% shock on Mortgage Backed Securities (1 asset class); and (iv) a 10\% shock on U.S. treasury securities, U.S agency securities, Securities issued by state and local governments (3 asset classes). The resulting aggregate vulnerability with real data and estimated with  CECAPM is shown in Figure \ref{fig:othershocks}. In all cases the CECAPM estimation tracks quite closely the AV obtained from the full knowledge of portfolios composition. We therefore conclude that our result is not due to the uniform shock assumption, but is more generically applicable.

\subsubsection{European Banking Authority Data}

We now show that our results hold also for different banking systems. To this end we investigate the  public dataset made available\footnote{  \url{http://www.eba.europa.eu/risk-analysis-and-data/eu-wide-stress-testing/2011/results}} by the European Banking Authority (EBA), after the 2011 stress tests conducted on the largest 90 European banks at the time. The data  consists of the exposures of each bank toward a set of 42 asset classes, as well as their book leverages. We defined the asset classes following precisely \cite{Greetal15} \footnote{The asset classes are: \enquote{sovereign debt of each of the 27 EU countries plus 10 others, commercial real estate, mortgages, corporate loans, small and medium enterprise loans, and retail revolving credit lines.}}.  Moreover the initial shock is \citep[following][]{Greetal15} \enquote{a 50\% write-off of all GIIPS debt}. As for the U.S. commercial banks, we compared the AV obtained from the full network data with that obtained using CECAPM.  The relative percentage bias of AV estimated using partial information is  $3.7 \%$. As a robustness check, we repeated the exercise for two different shocks:  a 10\%  write off of either all EU debt or all sovereign debt, including non E.U. countries. The percentage bias of the CECAPM estimation is  $3.6\%$ in the former case and $5.1 \%$ in the latter. Clearly also for this dataset, CECAPM gives a faithful estimation of AV, since the bias is around $3\%-5\%$, showing the robustness of the method.
 
\subsection{Assessing systemic risk for individual banks}\label{sec:performance}
 
We now test the performance of CECAPM in assessing systemic risk for individual banks. In order to assess the performance of an estimator that produces estimates $\widehat{S}_n$ and $\widehat{\textrm{IV}}_n$ of, respectively, systemicness and indirect vulnerability of the $n$-th bank 
in a given quarter we compute the relative error  as

\be\label{eq:perc_error}
s_{n} = \frac{\widehat{S}_n-S_n^{\star}}{S_n^{\star}},~~~~~~~~~~~~v_{n} = \frac{\widehat{\textrm{IV}}_n-\textrm{IV}^{\star}_n}{\textrm{IV}^{\star}_n}.
\ee

In each quarter we have from $N=6,500$ to $N=9,000$ values of relative errors for each metric. We find that the median of the distribution of relative errors for both metrics is quite constant and fluctuates around $\div-12\%$, while the interquantile range is between $-20\%$ and $15\%$ (see the figure in Appendix \ref{subsec:hetShockApp}). Thus   Cross Entropy tends to slightly underestimate single banks measures of systemic risk. In summary, the estimates of systemicness and indirect vulnerably for each single bank as provided by the CECAPM-implied matrix are satisfactorily accurate. Once more, the important message is that it is possible to achieve pretty accurate estimates of systemic risk metrics, at the aggregate or individual institution level, due to fire sales spillover {\it without} a full knowledge of the portfolio holdings of financial institutions. 

\section{Comparison with the Max-Entropy ensembles}\label{sec:mod_recns}

The  Cross-Entropy approach described in the previous section assumes
that the unknown matrix elements are those with minimal distance (as proxied by the cross-entropy function) from an a priori matrix. In this approach, the available economic information, which, in the specific case described above, consists of the quantities $A_n^{\star}$ and $C_{k}^{\star}$,
 is used to construct the guess following an economic intuition. 
 
A  different rationale constitutes the foundation of the Maximum-Entropy (ME) ensembles approach. This reconstruction method assumes, as standard in contexts of partial information, that the undisclosed 
quantities (in our case, the banks' portfolio holdings $X_{n,k}$) are random variables generated from an unknown statistical distribution.  The ME approach  amounts to take, among all the possible probability 
distributions, the one which maximizes the informational content of the economic constraints imposed during the maximization. This property follows directly from the definition of information as stated in the seminal paper by \cite{Sha48}. 

We define\footnote{In describing our ME approach, we largely follow the theoretical framework of \cite{Kol09}. } a  network statistical 
model as a set $\mathcal{X}$ of graphs, called {\it ensemble},  and a probability mass function $\mathbb{P}_{\boldsymbol{\vartheta}}$  indexed by a vector of model parameters $\boldsymbol{\vartheta}$. In formula it is expressed as the triplet
$$
\Set{\mathbb{P}_{\boldsymbol{\vartheta}} ,\mathcal{X}, \boldsymbol{\vartheta} \in \Xi},
$$
where $\Xi$ is a convex subset of $\mathbb{R}^{P}$, with $P$ the total number of parameters of the model. The set $\mathcal{X}$ is a countable set whose elements are graphs. 
In what follows, we will not distinguish between the graph and the 
associated matrix $\mathbf{X}$, i.e. the probability mass 
function is defined in the space of integer valued matrices. 
Moreover the probability mass function
$\mathbb{P}_{\boldsymbol{\vartheta}}: \mathcal{X}\ra\quadre{0,1}$
is such that $\sum_{\mathbf{X}\in\mathcal{X}}\mathbb{P}_{\boldsymbol{\vartheta}}\tonde{\mathbf{X}}=1$  and is allowed to depend on a vector of real parameters $\boldsymbol{\vartheta}\in\Xi$.
A model can be defined by explicitly giving the ensemble, the probability mass function along with the space $\Xi$ of the parameters, or by deriving
$\mathbb{P}_{\boldsymbol{\vartheta}}\quadre{\mathbf{X}}$ through the recurrent application of some generative mechanism or rule, either starting from an empty graph or by applying a randomization procedure to a reference graph.

In its most general formulation, the ME principle postulates to obtain the probability mass function $\mathbb{P}$
as that which maximizes the Shannon's entropy
$$
S=-\sum_{\mathbf{X}\in\mathcal{X}} \mathbb{P}\quadre{\mathbf{X}}\,\log\tonde{\mathbb{P}\quadre{\mathbf{X}}}
$$
subject to the normalization constraint
$$
\sum_{\mathbf{X}\in\mathcal{X}} \mathbb{P}\quadre{\mathbf{X}}=1,
$$
and, possibly, to further additional constraints.

There are two ways of imposing the constraints. In the first one, termed {\it microcanonical ensemble}, constraints are imposed {\it exactly}, i.e. only the graphs fulfilling all the constraints have non zero probability. In the second one, termed {\it canonical ensemble}, all the graphs have non zero probability and the constraints are satisfied on average over the distribution. There are advantages and disadvantages in both approaches. The microcanonical ensemble is economically more grounded, for example in the system under investigation here it implies that a given network realization has non zero probability only if each bank (asset class) has the same asset size (capitalization) as in real data. On the contrary, in the canonical ensemble also graphs where these values are very different from the real data might have non zero probability. Despite this undesirable property, we believe it is worth performing a comparison of the Cross-Entropy approach with the canonical ME for the following reasons:
\begin{enumerate}
\item Solving the problem in the microcanonical ensemble is typically extremely hard  or it requires extensive numerical simulations, randomizing the network by allowing moves that preserve all the constraints. On the contrary, canonical ensemble can often be obtained much more directly, as testified by their widespread use in Statistical Mechanics \citep[][]{huang}. Moreover when other constraints are added to the optimization problem, microcanonical ME (as well as Cross-Entropy) becomes intractable, thus limiting their practical use when regulators want to add additional knowledge on the system.
\item  The flexibility of canonical ME allows exploring the relative role of information set and constraints in network reconstruction. For example we will show below that, using the same information sets (the strength sequences) but {\it different} constraints can lead to very different performance in the estimation of systemic risk, indicating its main determinants.
\item  The excellent performance of CECAPM for systemic risk assessment calls for the construction of a network probability distribution which performs on average as CECAPM, but allows for scenario generation. The canonical ME ensemble we will introduce below (MECAPM) does exactly this job.
\item Last but not least, the application of canonical ME network ensembles in Economics and Finance is quite widespread, see, for example, \cite{bargigli2011random,squartini2011randomizing,fagiolo2013null,Masetal14,Saretal15,almog2017double}
\end{enumerate}

\subsection{Maximum Entropy ensembles}

We will consider three ME ensembles in this paper\footnote{All numerical routines, accompanied with an instruction manual, can be downloaded from \url{http://mathfinance.sns.it/network_reconstruction/}}. 
First, we propose a new max entropy ensemble which is based on the role of CAPM in the problem at hand. The probability mass function $\mathbb{P}$ is the solution of the optimization problem

\begin{equation}\label{eq:wcapm}
\begin{aligned}
& \underset{\mathbb{P}}{\text{max}}
& & - \sum_{\mathbf{X}\in\mathcal{X}} \mathbb{P}\quadre{\mathbf{X}}\,\log\tonde{\mathbb{P}\quadre{\mathbf{X}}} \\
& \text{s.t.} & & \sum_{\mathbf{X}\in\mathcal{X}}\mathbb{P}\quadre{\mathbf{X}}=1\\
& & & \mathbb{E}_{\mathcal{X}}\quadre{X_{n,k}}=\capm{n}{k},~n=1,...,N,~k=1,...,K.
\end{aligned}
\end{equation}

We call this model Maximum Entropy Capital Asset Pricing Model (shortened in MECAPM  henceforth). 
In Appendix \ref{app:mecapm_sol} we prove that the MECAPM has the unique solution 

\be\label{eq:pdf_MECAPM}
\mathbb{P}\quadre{\mathbf{X}} = \prod_{n=1}^N\prod_{k=1}^K\tonde{\frac{\capm{n}{k}}{1+\capm{n}{k}}}^{X_{n,k}}\,\frac{1}{1+\capm{n}{k}},
\ee

hence each single matrix entry $X_{n,k}$ is geometrically distributed with mean $\capm{n}{k}$. To understand the rationale behind this ensemble, we notice an interesting relation between the CECAPM and MECAPM estimation of AV under a uniform shock of asset returns. As shown in Appendix \ref{app:expform},   

\begin{equation}\label{eqapprox}
\E{S_n\tonde{\mathbf{X}}}=S_n\tonde{\mathbf{X}^{\textrm{CAPM}}}\left(1+\frac{A_n^*}{L^*}+\frac{\sum_{k=2}^KC_k}{\sum_{k=2}^KC_k^{\star2}}\right),
\end{equation}

where $\E{S_n\tonde{\mathbf{X}}}$ is the expected systemicness of bank $n$ under the MECAPM ensemble and $S_n\tonde{\mathbf{X}^{\textrm{CAPM}}}$ is the one according to the CECAPM. We notice that the former is larger than the latter, but the correction is small if $A_n^*\ll L^*$, since the last term in parenthesis is generally small. This result can also be used to compute systemicness and AV in the MECAPM ensemble {\it without} sampling but using the expression above. A similar result holds for indirect vulnerability (see Appendix \ref{app:expform} for details).

Since the other specifications of maximum entropy are quite popular in the 
literature of network reconstruction, for comparison purposes we take into considerations
two other ensembles, mainly inspired by the paper by \cite{Masetal14} and \cite{Saretal15}. Each of them is 
characterized by different constraints imposed on 
the maximization of the Shannon's entropy. 

In the first ensemble, termed Bipartite Weighted Configuration Model (BIPWCM), the 
constrained maximization is  

\begin{equation*}\label{eq:wcm}
\begin{aligned}
& \underset{\mathbb{P}}{\text{max}}
& & - \sum_{\mathbf{X}\in\mathcal{X}} \mathbb{P}\quadre{\mathbf{X}}\,\log\tonde{\mathbb{P}\quadre{\mathbf{X}}} \\
& \text{s.t.}
& & \sum_{\mathbf{X}\in\mathcal{X}} \mathbb{P}\quadre{\mathbf{X}}=1\\
& & &\mathbb{E}_{\mathcal{X}}\quadre{A_n}=A_n^{\star},~n=1,...,N,\\
& & & \mathbb{E}_{\mathcal{X}}\quadre{C_k}=C_k^{\star},~k=1,...,K.
\end{aligned}
\end{equation*}
Appendix \ref{subsec:bipwcm} reports the derivation and calibration of the ensemble. 
Note that BIPWCM imposes weaker constraints with respect to MECAPM, while exploiting the same 
information set, namely the strength sequences. 

Finally, we consider another (richer) statistical ensemble whose probability mass function, derived in Appendix \ref{subsec:bipecm}, corresponds in our bipartite framework to the enhanced configuration model of \cite{Masetal14}. This newly defined ensemble, that we address as Bipartite Enhanced Configuration Model (BIPECM), is obtained via Maximum Entropy imposing both the mean value of strengths (as in BIPWCM) and the mean value of degrees, that is the number of edges incident in each vertex. In other words, we reconstruct the matrix by assuming the knowledge of the number of assets in which each bank invests as well as the number of banks investing in each asset. Despite the fact that this information is typically not known, we consider this ensemble to show that even with an information set significantly larger than the one used in MECAPM it is very difficult to outperform it. Mathematically, the BIPECM is obtained by solving the optimization problem 

\begin{equation}\label{eq:ecm}
\begin{aligned}
& \underset{\mathbb{P}}{\text{max}}
& & - \sum_{\mathbf{X}\in\mathcal{X}} \mathbb{P}\quadre{\mathbf{X}}\,\log\tonde{\mathbb{P}\quadre{\mathbf{X}}} \\
& \text{s.t.}
& & \sum_{\mathbf{X}\in\mathcal{X}} \mathbb{P}\quadre{\mathbf{X}}=1\\
& & &\mathbb{E}_{\mathcal{X}}\quadre{A_n}=A_n^{\star},\\
& & & \mathbb{E}_{\mathcal{X}}\quadre{D_n^{\textrm{row}}}=D_n^{\textrm{row}^{\star}},~n=1,...,N,\\
& & & \mathbb{E}_{\mathcal{X}}\quadre{C_k}=C_k^{\star},\\
& & & \mathbb{E}_{\mathcal{X}}\quadre{D_k^{\textrm{col}}}=D_k^{\textrm{col}^{\star}},~k=1,...,K,
\end{aligned}
\end{equation}

where $D_n^{\textrm{row}}$ and $D_k^{\textrm{col}}$ are, respectively, 
the row and the column degree sequences (see Appendix \ref{subsec:bipecm} for more details).  
The peculiarity of BIPECM is the addition of the information 
on the degree sequences that is absent in both BIPWCM and MECAPM\footnote{One could consider another maximum entropy ensemble where the constraints are the same as in MECAPM plus the degree sequences. This is an enhanced MECAPM because additional information on the number of asset classes in each portfolio (and the number of banks investing in each asset class) is used. The optimization can be performed but the application on the US banks data shows no appreciable improvement in the systemic risk assessment with respect to the CECAPM (data available on request). For this reason and for the sake of simplicity in this paper we will not present results on this ensemble.}.  
Note that the three ensembles can be used not only for statistical inferences, but 
to produce estimates of any function defined on the network, which is the topic of the
next section. 

\subsection{Results}


\begin{figure}[t]
\begin{center}
\includegraphics[width=0.8\textwidth]{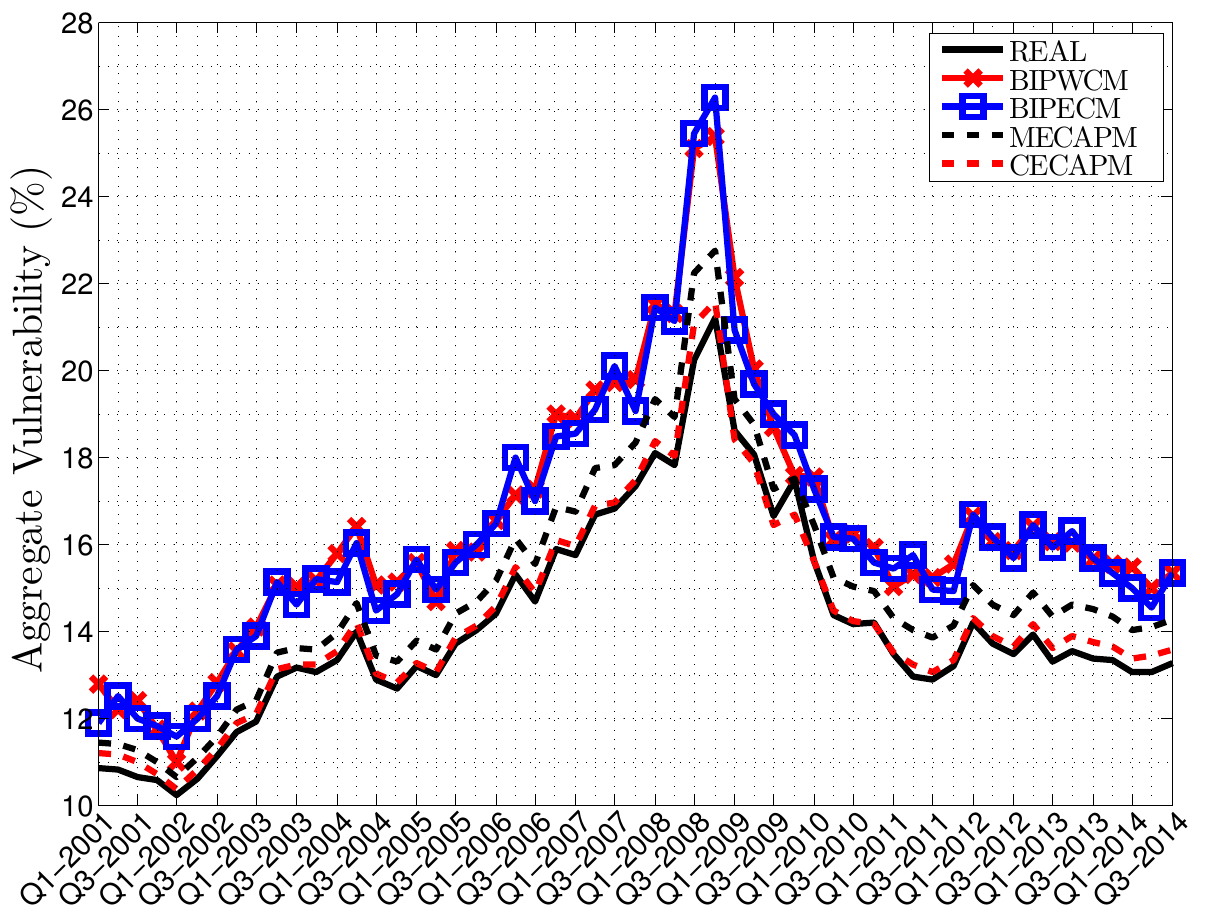}
\caption{This figure reports as a black continuous line the aggregated vulnerability, as 
defined by equation \eqref{eq:AV}, computed on the matrix $X_{n,k}^{\star}$ of portfolio 
holdings as provided by the FFIEC dataset of US commercial bank holdings, described in Section \ref{sec:data}. 
The other lines refer to the aggregate vulnerability reconstructed with the four entropic methods.}\label{fig:agg_sys}
\end{center}
\end{figure}

Figure \ref{fig:agg_sys} compares the true value of the aggregated vulnerability, obtained by using the real matrix of portfolio compositions, with those obtained with entropic methods. It is clear that all the methods track qualitatively well the temporal pattern of AV in the investigated period, but it is worth noticing that CECAPM has a very tiny bias, providing estimates of AVs in excellent agreement with the real one. As expected from the above argument, AV under MECAPM is always slightly larger than under CECAPM. Among the max entropy methods, MECAPM outperforms BIPWCM and  BIPECM. Since the information set required to derive the BIPECM is larger 
than that used for the MECAPM, this means that it is not the amount of information that matters, rather the way in which information
is conveyed in the reconstruction algorithm. Finally, notice that the true portfolio matrix is quite different from the matrices of CECAPM, MECAPM, and BIPWCM because in the former half of the matrix elements are zero while the latter models have adjacency matrices with all non vanishing elements. 

\begin{table} 
\begin{center}
 \begin{tabular}{rcc|cc}
& \multicolumn{2}{c|}{\rule{0pt}{15pt} 50\% GIIPS } & 10\%   E.U. Gov.  & 10\% All Gov. \\ 
&\multicolumn{2}{c|}{  $AV = 496.3\%$} &   $AV = 270.8\% $ &   $AV = 357 \% $\\
\hline
& \rule{0pt}{15pt} $\widehat{AV}$ & \% Bias	& \% Bias	& \% Bias \\
\hline
\rule{0pt}{12pt}CECAPM & 480.4  \%& 3.2\% &3.6 \%	& 5.1 \% \\ 
\rule{0pt}{12pt}BIPWCM &361.9\% & 27.1 \% & 28.6 \%	& 22.4 \% \\
\rule{0pt}{12pt}BIPECM &392.9 \%& 20.8\% &  20.6 \%	& 12.5 \% \\
\rule{0pt}{12pt}MECAPM & 436.7 \% &  12\% &  12.4 \%	& 4.1 \% \\
\hline
\end{tabular}
\caption{ Comparison, for the EBA data, between real and estimated values of the Aggregate Vulnerability ($AV$).  The three columns correspond to different shocks. Under the name of each shock we report the corresponding  real $AV$, computed from the complete knowledge of banks' portfolios. In the first column we report the estimated $AV$ and the percentage bias for the 4 different ensembles, resulting from a 50\% value loss of GIIPS sovereign debt, as considered in \cite{Greetal15}. In the second and third columns we report the percentage biases of estimated $AV$ for two alternative scenarios: a 10\% loss of value for either all the E.U. sovereign debt or the sovereign debt of all countries.}\label{tab:eba}   
\end{center}
\end{table} 

A similar comparative result holds by considering different shock scenario, as those studied in Section \ref{sec:cross_entropy} (see Fig. \ref{fig:othershocksAllModels} of the Appendix \ref{subsec:hetShockApp}) as well as for the European Banking Authority Data (see Table \ref{tab:eba}. Among the max entropy methods, MECAPM significantly outperforms BIPWCM and BIPECM in estimating the AV obtained with the full knowledge of the portfolio composition of banks. 

Finally we considered the assessment of systemic risk for individual banks. Figure \ref{fig:sysreconAllMethods} of the Appendix \ref{subsec:hetShockApp} shows  that for each quarter BIPWCM strongly underestimates individual bank systemicness and indirect vulnerability. The median relative error ranges roughly between $-60\%$ and $-70\%$ and the interquartile range is very far from zero. The estimator based on BIPECM (using the additional information on degrees) gives slightly better results, even if a strong underestimation is still present. The median relative error ranges roughly between  $-50\%$ and $-40\%$ and again the interquartile range is far from zero. On the contrary the estimator based on MECAPM (or CECAPM) performs much better. The median relative error never goes below $-20\%$ and almost always the interquartile range is centered around zero.\footnote{If instead we focus on the banks with higher systemicness or indirect vulnerability, the performances of the estimator based on MECAPM worsen. In particular,  for the quartile of banks with largest systemicness, the median percentage bias of the MECAPM estimator of systemicness is always between $-20\%$ and $-30\%$. Similarly, the median of the percentage bias in the estimation of indirect vulnerability via MECAPM is always between  $-20\%$ and $-35\%$. Nevertheless, the ranking among the three estimation methods remains unchanged.}
 
  

In summary, the estimates of systemicness and indirect vulnerably  for each single bank as provided by the CECAPM-implied matrix are almost identical to those obtained as the corresponding expected values on the MECAPM ensemble. Besides, they 
are satisfactorily accurate and surely more reliable than those provided by standard maximum entropy ensembles. 
Once more, the important message is that it is possible to achieve pretty accurate estimates of systemic risk metrics, at the aggregate or individual institution level, due to fire sales spillover {\it without} a full knowledge of the portfolio holdings of financial institutions.

\subsection{Monitoring and testing changes in systemicness}\label{sec:testing}

As another application of the ensembles of graphs obtained with the Maximum Entropy method, we consider here the problem of assessing whether the systemicness of a given bank (or of the whole system) has changed in a statistically significant way. In order to answer this question, it is necessary to have a null hypothesis and we propose to use network ensembles to this end. Since the MECAPM shows superior performances in 
estimating risk metrics, in this section we use it and we propose a possible application for
statistical validation. Our objective here is not to study all the banks and all the quarters, but only to show how the testing method can be implemented. 

\begin{figure}[t]
\thisfloatpagestyle{empty}
\begin{center}
\includegraphics[width=0.49\textwidth]{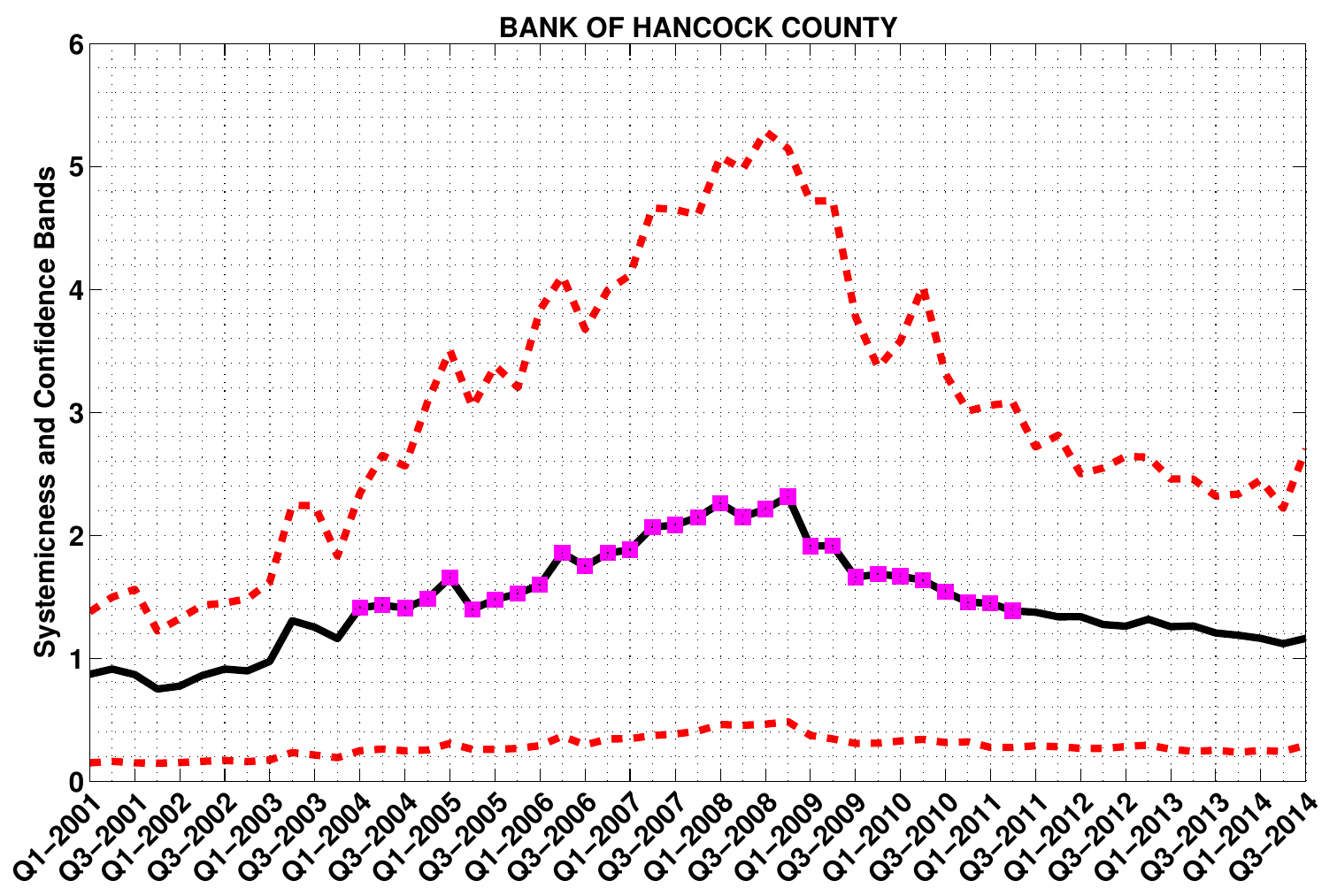}
\includegraphics[width=0.49\textwidth]{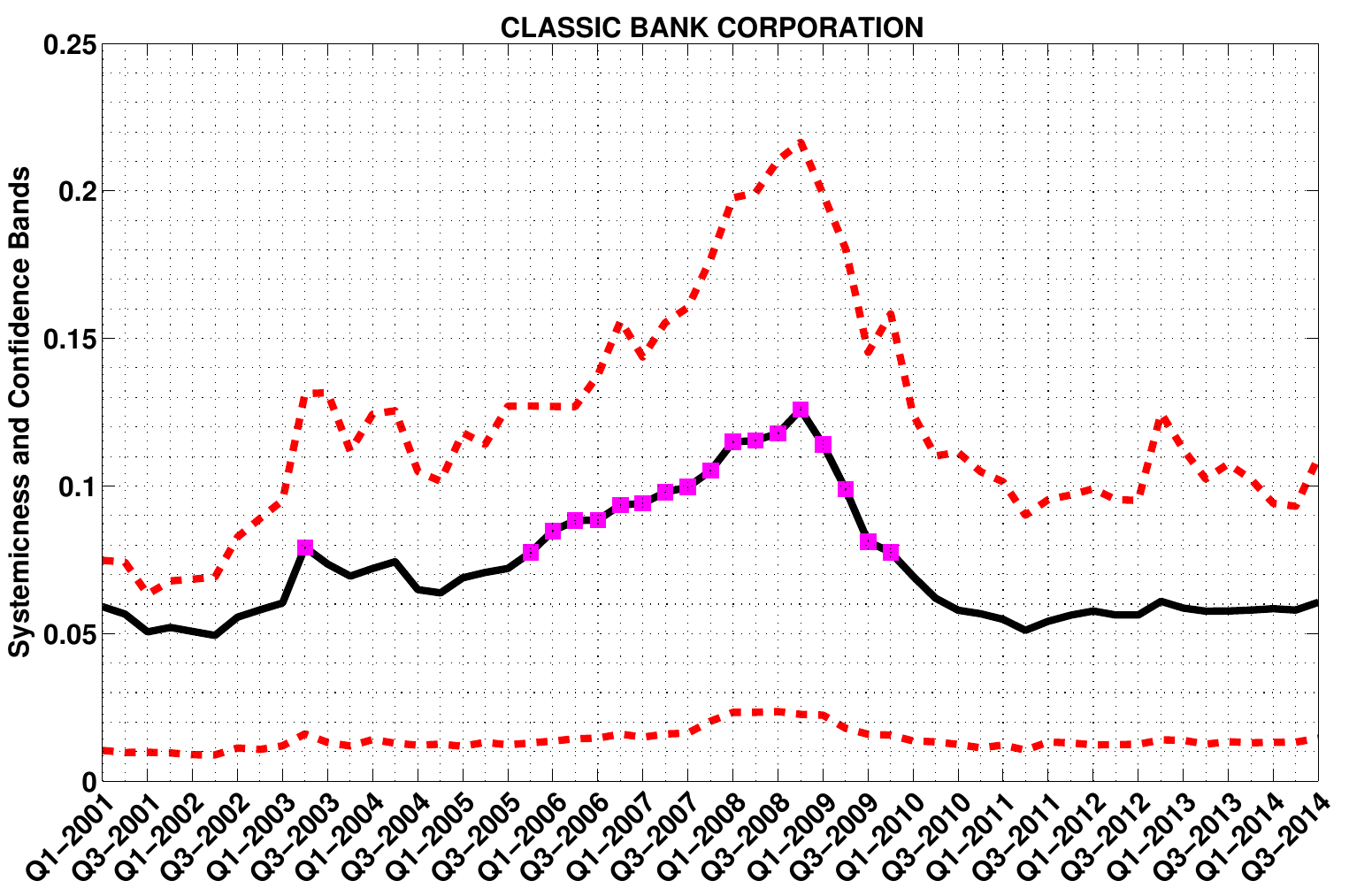}\\
\includegraphics[width=0.49\textwidth]{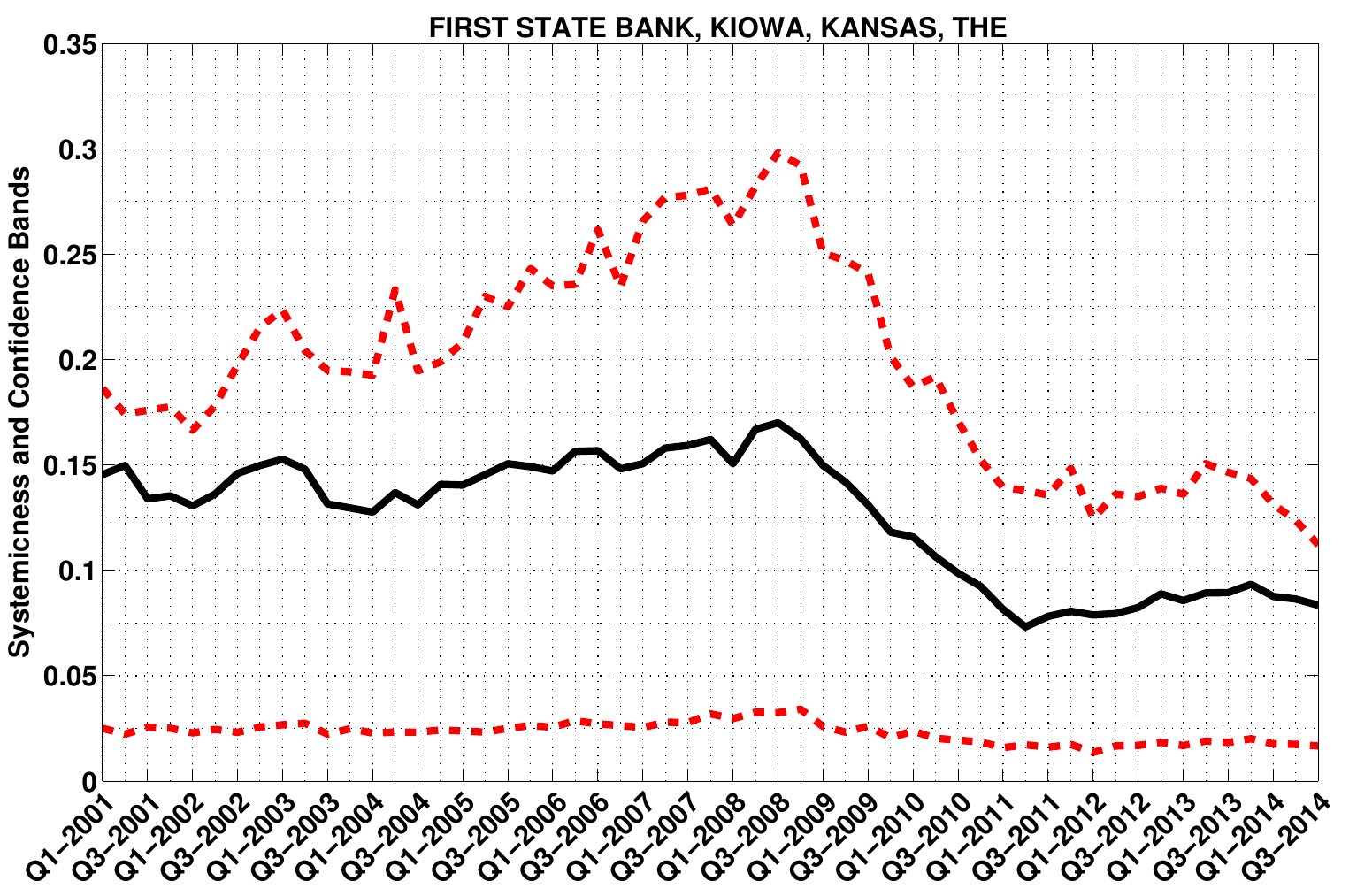}
\includegraphics[width=0.49\textwidth]{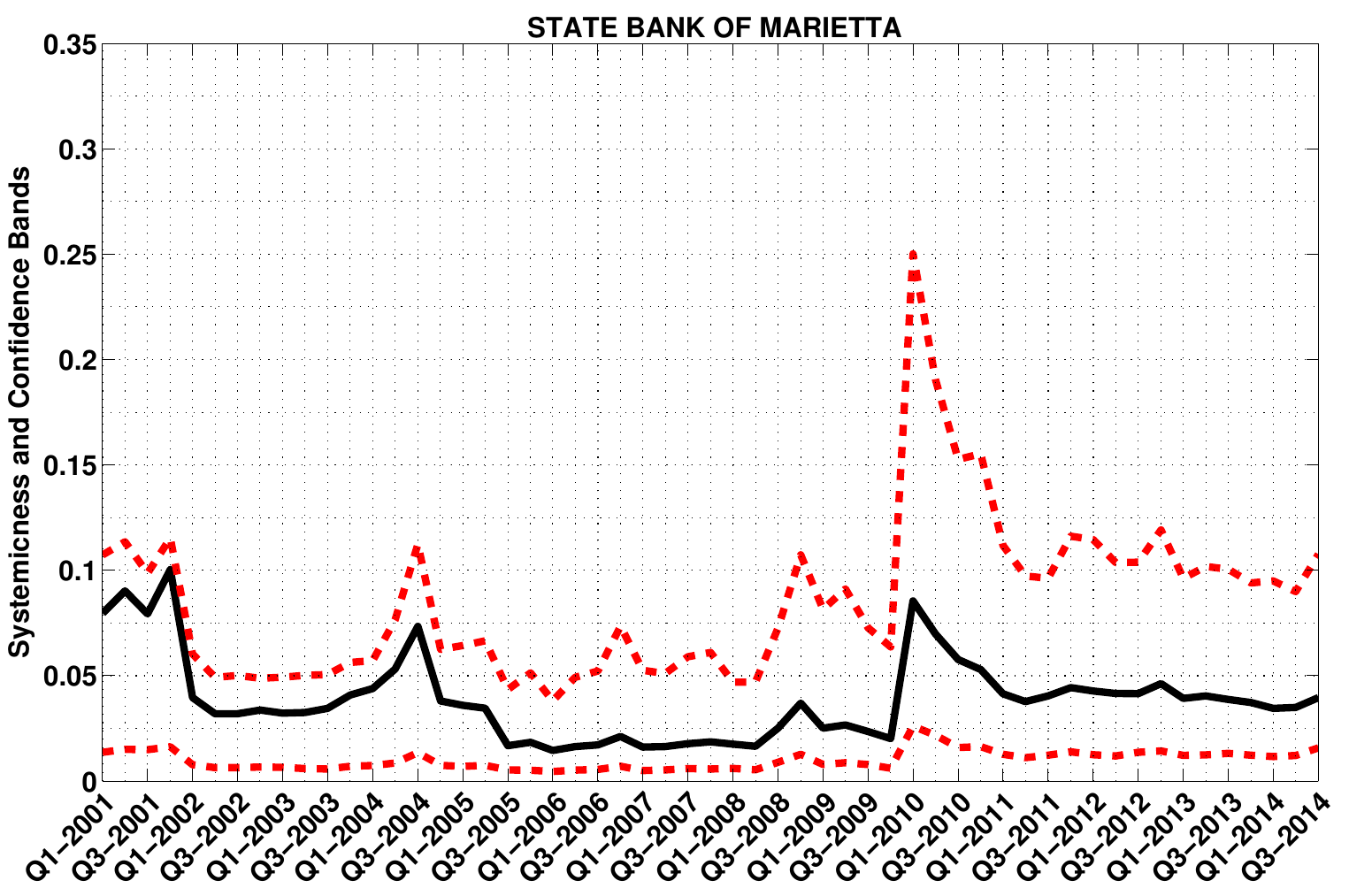}\\
\caption{ We report, for four selected banks, the true systemicness (thick dotted lines) and the 
$5\%$-$95\%$ confidence bands according to the MECAPM ensemble. A magenta square is added 
in every quarter in which the systemicness of the bank is above the $95\%$ confidence level
of the first quarter of $2001$.}\label{fig:sysconf}
\end{center}
\end{figure}

In particular, imagine a regulator who monitors a given bank, measuring its systemicness and searching for evidences of a significant increase. Having a given quarter as reference, the regulator can extract the distribution of bank's systemicness and, in the subsequent quarters, identify when the systemicness is outside a given confidence interval around the reference period. As a special case, we select four banks among the top fifty in the first quarter and that exist for the entire time period (i.e. they do not exit the dataset). 
For each quarter we compute the true bank systemicness and the $5\%$-$95\%$ confidence 
bands according to the MECAPM ensemble (see Figure \ref{fig:sysconf}).  
We then added a magenta square in each quarter when the true systemicness
is above the $95\%$ confidence band of the first quarter, used as reference. Hence, 
a magenta square is indicating a quarter when 
the systemicness of the bank is statistically larger (according to the MECAPM)
than at beginning of $2001$. We show two banks for which a statistically significant change in systemicness is observed (top row) and two for which no change is observed (bottom row). Notably, for the former case we find that 
the systemicness of the banks analyzed increased significantly 
much before the onset of the 2007-2008 financial crisis. This phenomenon 
persisted along the entire period of the crisis and vanished
not before the end of 2009. This suggests that network statistical models could
be of valuable help in the surveillance activity of central banks 
and other supervisory authorities as monitoring tools and in constructing early warning indicators.

\section{Conclusions}\label{sec:conc}
In this paper we focused on the problem of estimating metrics of systemic risk due to fire sale spillover in presence of limited information on the composition of portfolios of financial institutions. A full knowledge of the portfolio holdings of each institution in the economy is generally required to have a precise estimate of any risk metrics that, as those proposed by \cite{Greetal15}, is based on the mechanism of portfolio rebalancing through fire sales. Nevertheless, such a huge and detailed information may not be available, especially at frequency higher than quarterly, making the estimation of systemic risk quite difficult. In this paper we circumvent the problem by providing accurate estimates of systemic risk metrics that are based on a partial knowledge of the system, more precisely only on the sizes of balance sheets and the capitalization of assets (or asset classes), which are much easier to trace. In this respect, we have shown that the method of Cross-Entropy minimization does a very good job in estimating aggregate vulnerability and individual bank systemicness without requiring any knowledge of the underlying matrix of bank portfolio holdings.

Furthermore, we have compared the results with a Max Entropy ensembles. Specifically we have introduced a new ensemble (MECAPM), which reproduces, on average, the CECAPM and performs quite well in estimating systemicness and indirect vulnerability of single institutions, outperforming standard Max Entropy competitors. Moreover the estimation of systemic risk metrics could provide valuable information to any policy maker, but variations in systemicness and indirect vulnerability are difficult to interpret in absence of a statistical validation. For this reason, as a final contribution, we have proposed to use the Max Entropy ensemble to assess the statistical significance of systemic risk metrics. On a selection of banks of our dataset we documented that their systemicness significantly increased, with respect to the level observed at the beginning of the 2001, much before the onset of the 2007-2008 financial crisis. Even if deeper investigations are required in this direction, we believe that this approach could be easily implemented as an early warning indicator of systemic risk. 

Finally, we would like to comment again on the scope of the \cite{Greetal15}
model as well of our paper. As discussed in the main text, the considered methodology belongs to the classic static stress test approach. Only the portfolios and balance sheets at the time of the tested shock are used and no intertemporal dynamics is ever considered. This is a serious limitation, since financial distress and deleveraging might occur on longer periods and the bank's decision at a given quarter can depend, not only on the present price changes and portfolio composition, but also on past market state and banks' behavior. We believe that extending the \cite{Greetal15} approach to a dynamic stress test setting is a very interesting avenue for research both for academicians and for regulators.

\clearpage
\appendix

\section{Data Description and Dataset Creation}\label{app:appendix_data_description} 

\begin{figure}[htp!]
\begin{center}
\includegraphics[width=0.49\textwidth,height=0.3\textheight]{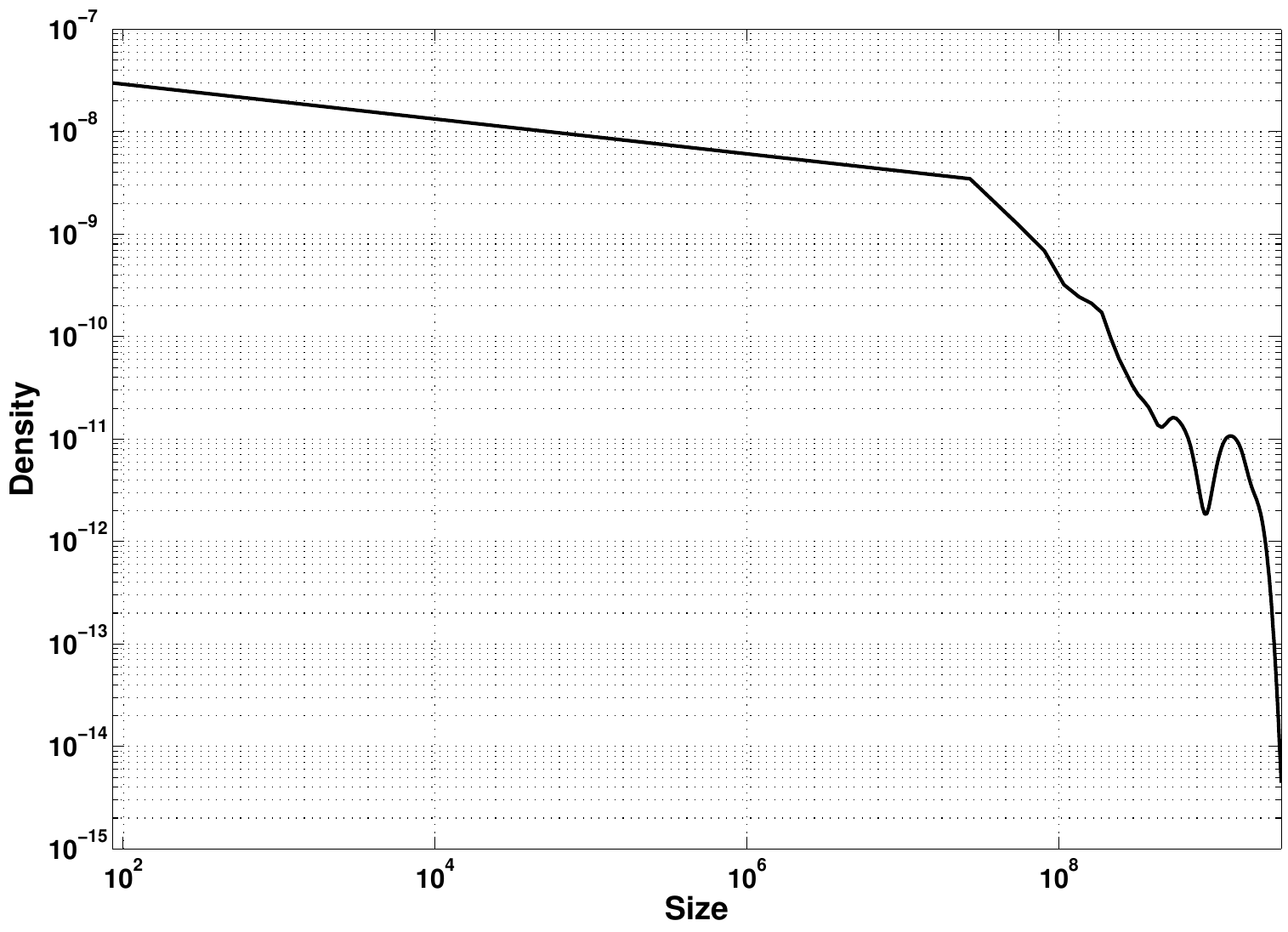}
\includegraphics[width=0.49\textwidth,height=0.3\textheight]{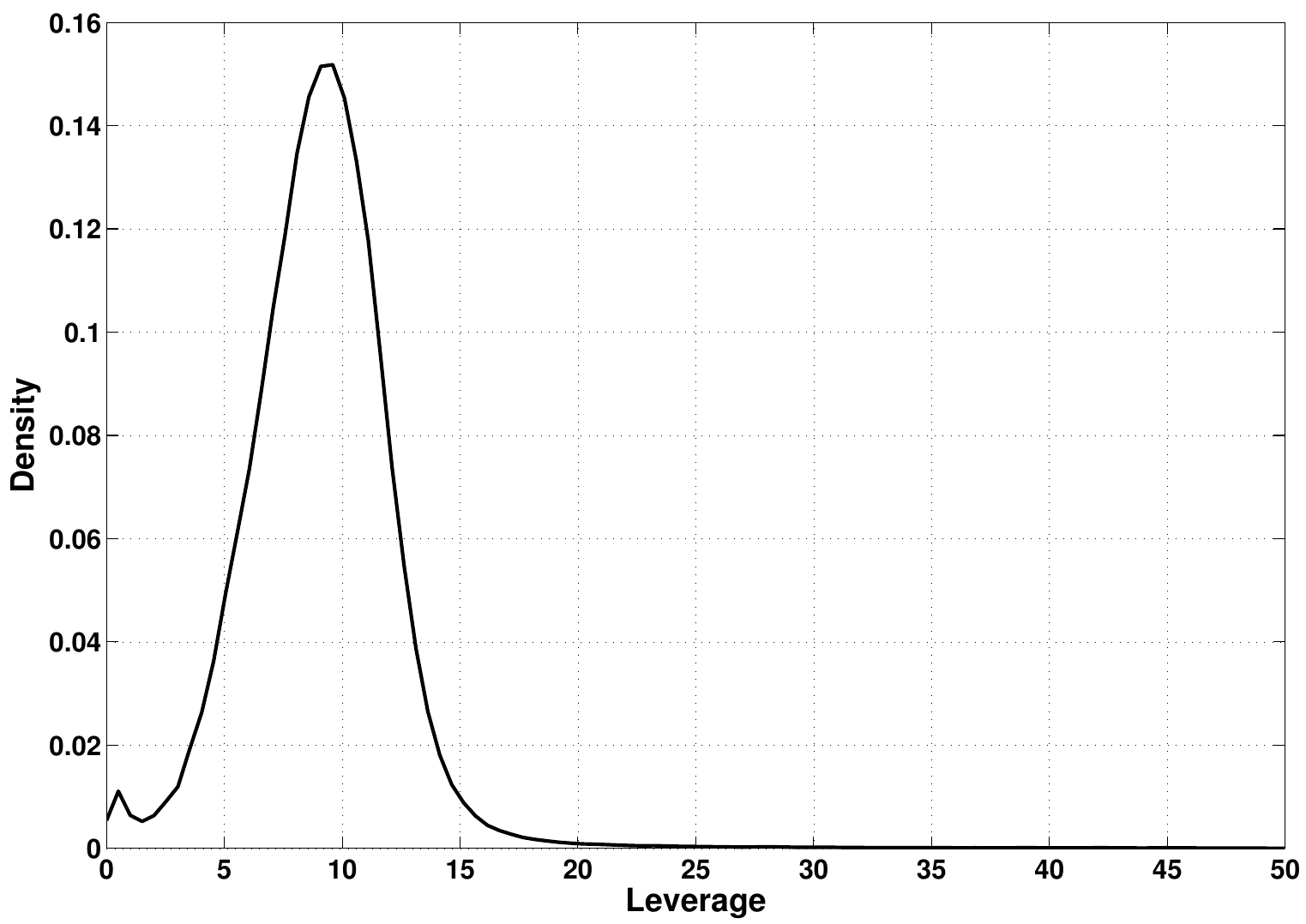}\\
\includegraphics[width=0.55\textwidth,height=0.3\textheight]{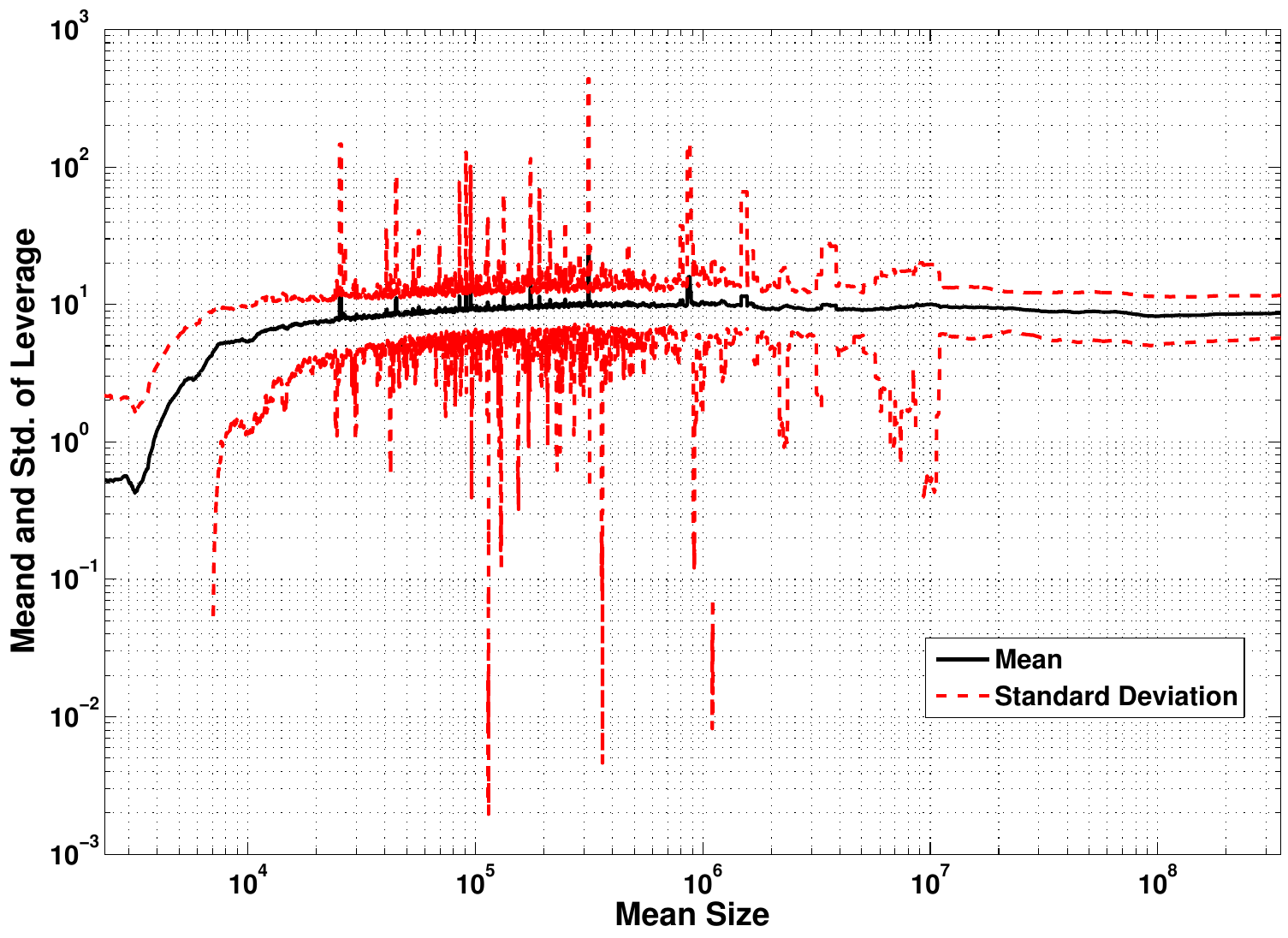}
\caption{\small{This Figure reports some descriptive features of the data analyzed. Top left panel plots,
on a log-log scale, the kernel density of bank sizes (defined as total assets in unit of $10^3\$$) while top right
is the kernel density of the bank leverages. Both densities are computed using all records pooled across the 
entire time span. For the sake of visualization, we put a cut-off of $50$ on the maximum leverage allowed, although 
leverages of more than $150$ are (rarely) observed. 
The bottom panel shows that there is no relation between leverage and size. The procedure
adopted to draw the plot is the following: all records
of bank size are sorted from the smallest to the largest one and a rolling window of $1000$ records is moved, 
with an incremental shift of $10$ records, from the first to the last. In each window we compute the 
mean leverage (black continuous line) and the standard deviation of leverage (red dotted line) of banks that fall
in the window. Mean and standard deviation are plotted as a function of the mean size in the window, which is reported in the horizontal axis.}}\label{fig:varie}
\end{center}
\end{figure}

This appendix provides some descriptive features of the data along with 
the method adopted to build the $20$ asset classes of the bank-asset network analyzed in the paper.
The left panel (first row) of Figure \ref{fig:varie} reports, on a log-log scale, the kernel density of the 
bank sizes (i.e. the total amount of assets detained by the bank) pooled across 
all quarters. It is evident that bank sizes are quite heterogeneous. The right panel (first row) of Figure \ref{fig:varie}
reports the density  of the bank leverages $B_n$ pooled across all quarters. In this case
we observe a much less heterogeneous distribution, with most banks showing a 
leverage around $10$.  Finally, the second row of Figure \ref{fig:varie} 
reports the relation between size and leverage. The plot is achieved 
by sorting all records of bank size from the smallest to the largest and then 
applying a moving-window procedure. 
As expected from the density plots, there is no relation between leverage and bank size, having 
most bank a leverage of $10$ and a highly heterogeneous size.

Concerning the formation of the asset classes used in the main text, we provide 
in what follows details on how they have been created. As mentioned in the main text, 
the focus of the paper is on commercial banks, whose precise definition is given by the FFIEC as 
\textit{[...] every national bank, state member bank, 
 insured state nonmember bank, and savings association is required 
 to file a consolidated Call Report normally as of the close of business 
 on the last calendar day of each calendar quarter, i.e., the report date. The specific reporting 
 requirements depend upon the size of the bank 
 and whether it has any \enquote{foreign} offices [...]}.
This is the set of institutions that is referred as Commercial Banks throughout all the paper.

Forms FFIEC031 and FFIEC041  
are dedicated to, respectively, banks with only domestic offices and 
banks with domestic and foreign offices. However, in both forms, it is adopted 
the same coding system. More specifically there are only two types of 
codes, RCON and RCFD, which are followed by a four digits alphanumerical code. The alphanumerical
code identifies the budget item, for example $2170$ refers to total assets of the bank. The prefix
RCON is used for financial items relative to domestic offices, while RCFD encompasses both 
domestic and foreign offices. Hence $\textrm{RCON}2170$ is the code for the total assets of the bank 
detained in U.S. offices, while $\textrm{RCFD}2170$ is relative to the sum of total assets
detained in U.S. plus offices abroad. Of course, for banks that fill the FFIEC031 the two 
codes $\textrm{RCON}$ and $\textrm{RCFD}$ report the same value if they have the same alphanumerical code. 

Table \ref{tab:ass_com} documents the detailed composition of each asset class. 
For each asset class (first column) we report the composition in terms of 
FFEIC items in the third column and a short name given to the asset class in the second one. Such abbreviation
is needed since some asset class, e.g. \enquote{loans to consumers in foreign offices}, are assembled subtracting 
from the FFIEC codes some previously defined asset classes. There is a one-to-one correspondence between asset classes
and variable names, a part for the case of \enquote{loans secured by real estates in domestic offices}, which
is computed as the sum of five variables, from \enquote{construction loans} to \enquote{non farm, non residential}.
The composition of the FFEIC formula reported
in the third column may vary during time, hence we report in bold the period of validity 
of the formula adopted. In this respect, note that the date $\textbf{12/99}$ refers to the last available quarter, that is the 
third quarter of $2014$. In reporting the FFEIC formula, we adopt 
the convention that the prefix is omitted whenever $\textrm{RCON}$ is used solely for banks with only domestic offices and 
$\textrm{RCFD}$ solely  for those that have at least on office abroad. 
On the contrary, when the prefix is specified, it means that only the code with that particular 
prefix is being been used. For example the code $\textrm{RCON}3532$ is used only in its domestic version,
hence we do not use $\textrm{RCFD}3532$ for banks with offices abroad.

\newgeometry{left=1.5cm,right=1.5cm}
\restoregeometry

\newgeometry{left=2cm,right=2cm}
\begin{table}
\thisfloatpagestyle{empty}
\vspace{-2.5cm}
\renewcommand\arraystretch{1.2}
\begin{center}
\caption{Composition of Asset Classes}
\resizebox{0.98\textwidth}{!} {
\begin{tabular}{lll}
  & & \\
Asset Class & Variable Name & FFIEC Formula  \\
\hline\hline
                                                     Total assets & tot\_ass & \textbf{03/01-12/99}: 2170+2123+3123  \\
\\
				                      Equity & equity  &  \parbox{9cm}{\textbf{03/01-03/09}: 3210+3000 \newline \textbf{03/09-12/99}:  G105} \\ 				                    
\\
   			        \parbox{9cm}{Cash and balances  due from depository institutions} &  cahab & \textbf{03/01-12/99}:  0081+0071 \\
\\
							 \parbox{9cm}{U.S. treasury securities} &  ust\_sec & \textbf{03/01-12/99}: 0211+1287+RCON3531 \\	
\\
							  \parbox{9cm}{U.S agency securities} & agency\_sec & \textbf{03/11-12/99}: 1289+1294+1293+1298+RCON3532  \\
\\	
							  \parbox{9cm}{Securities issued by state and local governments} & state\_sec  & \textbf{03/01-12/99}: 8496+8499+RCON3533	\\							
\\
							  \parbox{9cm}{Mortgage backed securities} & mbs &\textbf{03/01-03/09}: 1698+1702+1703+1707+1709+1713+1714+1717+1718+1732+\\
							 & & 1733+1736+RCON3534+RCON3535+RCON3536.\\
							 & &  \textbf{06/09-12/10}: G300+G303+G304+G307+G308+G311+G312+G315+G316+\\
							 & & \parbox{7cm}{G319+G320+G323+G324+G327+G328+G331+RCONG379+RCONG380+}  \\				
							  & & \parbox{7cm}{RCONG381+RCONG382}\\			
							 & & \textbf{03/11-12/99}: G300+G303+G304+G307+G308+G311+G312+G315+G316+\\
							 & & G319+G320+G323+K142+K146+K145+\\
							 & & K149+K150+K154+K153+K157+\\
							 & & \parbox{7cm}{RCONG379+RCONG380+RCONG381+RCONK197+RCONK198} \\							
\\
							  \parbox{9cm}{Asset backed securities} & abs & \parbox{9cm}{\textbf{03/01-12/05}: B838+B841+B842+B845+B846+B849+B850+B853+B854+}\\
							 & & \parbox{7cm}{B857+B858+B861 \newline 03/06-03/09 C026+C027 \newline \textbf{06/09-12/99}: C026+C027+G336+G340+G344+G339+G343+G347} \\							
\\
							  \parbox{9cm}{Other domestic debt securities} & dom\_debt\_oth\_sec  & \textbf{03/01-12/99}: 1737+1741 \\	
\\
							  \parbox{9cm}{Foreign debt securities} &  for\_debt\_sec & \textbf{03/01-12/99}: 1742+1746 \\			
\\
							  \parbox{9cm}{Residual securities} & res\_sec & \textbf{03/01-12/99}: A511 \\
\\
							 \parbox{9cm}{Futures, forwards sold and securities purchased under the agreement to resell (asset)} & ffrepo\_ass &  \parbox{9cm}{\textbf{03/01-12/01}: 1350 \newline \textbf{03/02-12/99}: RCONB987+B989} \\							
\\
							\multirow{5}*{\parbox{9cm}{Loans secured by real estates in domestic offices} }  &Construction loans &  \parbox{9cm}{\textbf{03/01-12/07}: RCON1415 \newline  \textbf{03/08-12/99}: RCONF158+RCONF159}\\							
\\
							 &  Secured by farmland & \textbf{03/01-12/99}: RCON1420\\					
\\
							  & 1-4 Family real estate & \textbf{03/01-12/99}: RCON5367+RCON5368+RCON1797\\
\\						
							  & Multifamily property loans  & \textbf{03/01-12/99}: RCON1460 \\
\\
							  & Non farm, non residential &  \parbox{9cm}{\textbf{03/01-12/07}: RCON1480 \newline \textbf{03/08-12/99}: RCONF160+RCONF161}\\							
\\
							  \parbox{9cm}{Loans secured by real estate in foreign offices} &  ln\_re\_for &  \parbox{9cm}{\textbf{03/01-12/99}: (if present) RCFD1410 - ln\_const - ln\_farm - ln\_rre - ln\_multi - ln\_nfnr, }\\ 
							 & & \textbf{3/01-12/99}: (otherwise)  zero\\
\\							
							  \parbox{9cm}{Commercial and industrial loans in domestic offices} & ln\_ci\_dom & \textbf{03/01-12/99}: RCON1766 \\							
\\							
							  \parbox{9cm}{Commercial and industrial loans in foreign offices} &  ln\_ci\_for & \textbf{03/01-12/99}: (if present) RFCD1763+RFCD1764 - RCON1766,  \\
							 & & \textbf{03/01-12/99}: (otherwise) zero \\
\\							
							  \parbox{9cm}{Loans to consumers in domestic offices} & ln\_cons\_dom  & \parbox{9cm}{\textbf{03/01-12/10}: RCON2011+RCONB538+RCONB539}\\
							 & &  \textbf{03/11-12/99}: +RCONB538+RCONB539+RCONK137+RCONK207\\								
\\							
							  \parbox{9cm}{Loans to consumers in foreign offices} & ln\_cons\_for  &\textbf{03/01-12/10}:  (if present) RCFD2011+ RCFDB538+ \\
							 & &  RCFDB539  - ln\_cons\_dom, (otherwise) zero\\
							 & & \textbf{03/11-12/99} (if present) RCFDB538+RCFDB539+ \\
							 & & RCFDK137+RCFDK207-ln\_cons\_dom, (otherwise) zero\\															
\\	
						         \parbox{9cm}{Loans to depository institutions and acceptances of other banks} & ln\_dep\_inst\_banks &  \textbf{03/01-12/99}: (if present) RCFDB532+RCFDB533+RCFDB534+\\
							 & & RCFDB536+RCFDB537, (otherwise) RCON1288 \\							
\\
							  \parbox{9cm}{other loans} & oth\_loans  & \textbf{03/01-12/99}: 2122+2123-ln\_const-ln\_farm-ln\_rre-ln\_multi-ln\_nfnr-\\
							 & & ln\_re\_for-ln\_ci\_dom-ln\_ci\_for-ln\_cons\_dom-ln\_cons\_for-ln\_dep\_inst\_banks\\
\\															
							  \parbox{9cm}{Equity securities that do not have readily determinable fair value} & equ\_sec\_nondet & \textbf{03/01-12/99}: 1752 \\								
\\					
							  \parbox{9cm}{other assets} & oth\_ass  & \textbf{03/01-12/99}: tot\_ass -  all preceding assets\\	
							\hline													
\end{tabular}
}
\vspace{0.1cm}
     \label{tab:ass_com} 
\end{center} 
\end{table}
\restoregeometry

\section{Max Entropy Ensembles}\label{app:bip_sol}
In this appendix we provide the details of the derivation of the probability mass functions
for the MECAPM, BIPWCM and BIPECM ensembles. In what follows we indicate with $S\tonde{\mathbb{P}}\quadre{\mathbf{X}}$ the entropy function  
$$
S\tonde{\mathbb{P}\quadre{\mathbf{X}}}=- \sum_{\mathbf{X}\in\mathcal{X}} \mathbb{P}\quadre{\mathbf{X}}\,\log\tonde{\mathbb{P}\quadre{\mathbf{X}}},
$$  
 and with $\phi_i\tonde{\mathbf{X}}$, $i=1,...,I$, a set of functions of the adjacency matrix, that describe the economic information available and that, in the constrained maximization, are required to have fixed expected values $\phi_i^\star$. Examples of $\phi_i\tonde{\mathbf{X}}$ are $\phi_{nk} = X_{nk}$ and $\phi_{n} = \sum_k X_{nk}$ or $\phi_{k} =\sum_n X_{nk}$.  The ME approach consists in defining the probability mass function which solves the constrained optimization problem
\be\label{eq:max_entr_opt_gen_constr}
\begin{aligned} 
& \underset{\mathbb{P}\tonde{\mathbf{X}}}{\text{max}}
& & S\tonde{\mathbb{P}\quadre{\mathbf{X}}} \\
& \text{s.t.}
& & \sum_{\mathbf{X}\in\mathcal{X}}\mathbb{P}\tonde{\mathbf{X}}=1,\\
& & & \mathbb{E}_{\mathcal{X}}\quadre{\phi_i }= \phi_i^\star,~ i =1,...,I.
\end{aligned}
\ee
If each matrix element can take only a finite set of values, i.e. the space $\mathcal{X} $ is finite, the optimization problem is easily solved, at least formally, using Lagrange multipliers.
The Lagrangian associated to the problem is written as
\ba
\mathcal{L} &=& S\tonde{\mathbb{P}\quadre{\mathbf{X}}}  +\alpha \left( 1 -\sum_{\mathbf{X} \in \mathcal{X}}\mathbb{P}\tonde{\mathbf{X}}   \right)+ \sum_{i=1}^{I}\vartheta_{i} \left( \phi_i^{\star} -  \sum_{\mathbf{X} \in \mathcal{X}} \mathbb{P}\tonde{\mathbf{X}} \phi_i\tonde{\mathbf{X}}  \right),\nonumber
\ea 
where $\alpha$ and $\vartheta_{i}$ are Lagrange multipliers. Taking the first derivative w.r.t. $\mathbb{P}\tonde{ x }\equiv\mathbb{P}\tonde{X  = x }$, where $x$ indicates an element of $\mathcal{X}$, we get\footnote{We stress that we are considering the derivative w.r.t. the probability of each of the (finite) possible realizations of $\mathbf{X}\in \mathcal{X}$.}	
$$
\frac{\partial \mathcal{L}}{\partial\mathbb{P}\tonde{x} } = -  \log(\mathbb{P}\tonde{x} ) - 1 - \alpha - \sum_{i=1}^{I}\vartheta_{i}\,\phi_{i} = 0, 
$$ 
whose solution is
\be\label{eq:max_entr_pdf}
\mathbb{P}_{\boldsymbol{\vartheta}}\tonde{\mathbf{X}}   = \frac{e^{-\sum_{i=1}^{I} \vartheta_{i}\phi_i\tonde{\mathbf{X}} }}{Z_{\boldsymbol{\vartheta}}},
\ee
where $\vartheta$ indicates the set of all Lagrange multipliers, 
 and  $Z_{\boldsymbol{\vartheta}}$ is a normalizing factor given by
\be\label{eq:partition_emcapm}
Z_{\boldsymbol{\vartheta}}  =  \sum_{\mathbf{X} \in \mathcal{X}}  e^{-\sum_{i=1}^{I} \vartheta_{i}\phi_i\tonde{\mathbf{X}}}.
\ee
Given a set of constraints $\phi_i^\star$ the corresponding Lagrange multipliers $\boldsymbol{\vartheta}^\star$ are obtained  as the (unique\footnote{The uniqueness of the solution is well-known in maximum entropy literature.}) solution  to
$$
\sum_{\mathbf{X} \in \mathcal{X}} \mathbb{P}_{\boldsymbol{\vartheta}^\star}\tonde{\mathbf{X}} \phi_i\tonde{\mathbf{X}}  = \phi_i^\star ,~ i =1,...,I.
$$
The latter parameters are determined, either analytically or numerically, from the economic information codified in the values $\phi_i^\star$. In the text and in the rest of this Appendix, we omit the dependency of $\mathbb{P}$ on $\boldsymbol{\vartheta}$.

 Following the literature cited in Section \ref{sec:mod_recns}, we consider models that allow $X_{nk}$ to assume any positive integer values\footnote{In considering this case we follow the standard approach in the literature. Requiring additional constraints on the set of values that each $X_{nk}$ can assume, would complicate the analytical computations that follow in this appendix, and result in more involved numerical estimation of the model.}, in which case the Lagrange multipliers derivation of the probability mass function is only heuristic. Nevertheless, it is well known \citep[see for example][]{campbell1970equivalence,barndorff2014information} that, even in the infinite case, the maximum entropy probability mass function is the one in \eqref{eq:max_entr_pdf} . In fact, it can be easily shown that every other probability mass function, satisfying the constraints in \eqref{eq:max_entr_opt_gen_constr}, has lower entropy than the one obtained with Lagrange multipliers. 
 
\subsection{Maximum Entropy Capital Asset Pricing Model}\label{app:mecapm_sol}
 Considering the maximum entropy optimization problem with constraint functions $\phi_{nk} = X_{nk}$, and $\lambda_{n,k}$ as Lagrange multipliers, we obtain the normalizing factor
\ba\label{eq:partition_emcapm}
Z_{\boldsymbol{\vartheta}} & = & \sum_{\mathbf{X} \in \mathcal{X}}  e^{-\sum_{n=1}^{N}\sum_{k=1}^{K}\lambda_{n,k}X_{n,k}}\accapo
                           & = & \sum_{\mathbf{X} \in \mathcal{X}}\prod_{n=1}^{N}\prod_{k=1}^{K}e^{-\lambda_{n,k}X_{n,k}}\accapo
                           & = & \prod_{n=1}^{N}\prod_{k=1}^{K}\sum_{X_{n,k}=0}^{\infty}e^{-\lambda_{n,k}X}\accapo
                           & = & \prod_{n=1}^{N}\prod_{k=1}^{K}\frac{1}{1-e^{-\lambda_{n,k}}}.
\ea
Hence 
$$
\mathbb{P}\tonde{\mathbf{X}} = \prod_{n=1}^{N}\prod_{k=1}^{K} \frac{e^{-\lambda_{n,k}\,X_{n,k}}}{1-e^{-\lambda_{n,k}}}.
$$
Note that the {\it partition function}  $Z_{\boldsymbol{\vartheta}}$ in \eqref{eq:partition_emcapm}
is such that
\ba\label{eq:derivlog}
\frac{\partial\log\tonde{Z_{\boldsymbol{\vartheta}} } }{\partial \lambda_{n,k}} &=&-\mathbb{E}_{\mathcal{X}}\quadre{X_{n,k}}.
\ea
Hence,  imposing the CAPM structure as required in \eqref{eq:wcapm},  the Lagrange multipliers are determined by
$$
-\frac{\partial\log Z_{\boldsymbol{\vartheta}}}{\partial \lambda_{n,k}} = \frac{A_n^{\star}\,C_k^{\star}}{L^{\star}},
$$
which gives  the probability mass function for the MECAPM 
\be\label{eq:ens_geom}
\mathbb{P} = \prod_{n=1}^{N}\prod_{k=1}^{K}\tonde{\frac{X^{\textrm{CAPM}}_{n,k}}{1+X^{\textrm{CAPM}}_{n,k}}}^{X_{n,k}}\,\tonde{\frac{1}{1+X^{\textrm{CAPM}}_{n,k}}},
\ee
where
$$
X^{\textrm{CAPM}}_{n,k} = \frac{A_n^{\star}\,C_{k}^{\star}}{L^{\star}}.
$$

\subsection{Bipartite Weighted Configuration Model.}\label{subsec:bipwcm}
 When we want to use as economic information the total asset size of each bank and the total capitalization of each asset, without imposing the MECAPM structure, we need one Lagrange multiplier $\lambda_n$ for each $\phi_{n} = \sum_k X_{nk}$, and one Lagrange multiplier $\eta_k$ for each $\phi_{k} = \sum_n X_{nk}$.  
We can go on with computation by explicitly writing the expression of $A_n\tonde{\mathbf{X}}$ and $C_k\tonde{\mathbf{X}}$ 
in terms of the elements of the matrix $\mathbf{X}$, obtaining 
\ba
Z_{\boldsymbol{\vartheta}} &=&   \sum_{\mathbf{X} \in \mathcal{X}}\prod_{n=1}^{N}\prod_{k=1}^K e^{-X_{n,k}\,\tonde{\lambda_n\,+\eta_k}}\accapo
                                           &=&  \sum_{X_{1,1}=0}^{\infty}\cdots\sum_{X_{1,K}=0}^{\infty}\cdots\sum_{X_{n,1}=0}^{\infty}\cdots\sum_{X_{n,K}=0}^{\infty}\prod_{n=1}^{N}\prod_{k=1}^K e^{-X_{n,k}\,\tonde{\lambda_n\,+\eta_k}}\accapo
                                           &=& \prod_{n=1}^{N}\prod_{k=1}^K \sum_{X_{n,k}=0}^{\infty} e^{-X_{n,k}\,\tonde{\lambda_n\,+\eta_k}}\accapo
 				         &=& \prod_{n=1}^{N}\prod_{k=1}^K \frac{1}{1-e^{-\tonde{\lambda_n\,+\eta_k}}}= \prod_{n=1}^{N}\prod_{k=1}^K \frac{1}{1-\varphi_n\,\xi_k},\nonumber
\ea
where $\varphi_n=e^{-\lambda_n}$ and $\xi_k=e^{-\eta_k}$, whence 
\ba\label{eq:pdf_BIPWCM}
\mathbb{P}\tonde{\mathbf{X}}&=& \frac{\prod_{n=1}^{N}\prod_{k=1}^K  e^{-X_{n,k}\,\tonde{\lambda_n\,+\eta_k} }}{\prod_{n=1}^{N}\prod_{k=1}^K \frac{1}{1-\varphi_n\,\xi_k} }\accapo
                                                                             &=&  \prod_{n=1}^{N}\prod_{k=1}^K  \tonde{\varphi_n\,\xi_k}^{X_{n,k}}\,\tonde{1-\varphi_n\,\xi_k},
\ea
The value of the Lagrange multipliers are determined by imposing that 
the expected value of $A_n\tonde{\mathbf{X}}$ and $C_{k}\tonde{\mathbf{X}}$ on the ensemble $\mathcal{X}$ are equal to, 
respectively, $A_n^{\star}$ and $C_k^{\star}$. As for the MECAPM case, note that $Z_{\boldsymbol{\vartheta}}$ is such that
$$
\mathbb{E}_{\mathcal{X}}\quadre{A_n} = -\frac{\partial\log\tonde{Z_{\boldsymbol{\vartheta}} } }{\partial \lambda_n},
$$
and similarly
$$
\mathbb{E}_{\mathcal{X}}\quadre{C_k} = -\frac{\partial\log\tonde{Z_{\boldsymbol{\vartheta}} } }{\partial \eta_k}.
$$
Hence we can compute $\mathbb{E}_{\mathcal{X}}\quadre{A_n}$ and $\mathbb{E}_{\mathcal{X}}\quadre{C_k}$ 
explicitly as a function of the Lagrange multipliers, that is
\ba
\mathbb{E}_{\mathcal{X}}\quadre{A_n} &=& \frac{\partial}{\partial \lambda_n}\sum_{n=1}^{N}\sum_{k=1}^K \log\tonde{1-e^{-\tonde{\lambda_n\,+\eta_k}}}\accapo
                                                               &=&  \sum_{k=1}^K\frac{e^{-\tonde{\lambda_n+\eta_k}}}{1-e^{-\tonde{\lambda_n\,+\eta_k}}},\accapo
\mathbb{E}_{\mathcal{X}}\quadre{C_k} &=& \frac{\partial}{\partial \eta_k}\sum_{n=1}^{N}\sum_{k=1}^K \log\tonde{1-e^{-\tonde{\lambda_n\,+\eta_k}}}\accapo
                                                               &=&  \sum_{n=1}^N\frac{e^{-\tonde{\lambda_n+\eta_k}}}{1-e^{-\tonde{\lambda_n\,+\eta_k}}}.\nonumber                                                               
\ea
Therefore the Lagrange multipliers are determined by numerically solving the non-linear system of equations
\be\label{eq:nnlinsys}
\left\{
\begin{array}{lll}
\sum_{k=1}^K\frac{\varphi_n\,\xi_k}{1-\varphi_n\,\xi_k} & = & A_n^{\star},~n=1,...,n,\\
 & & \\
 \sum_{n=1}^N\frac{\varphi_n\,\xi_k}{1-\varphi_n\,\xi_k} & = & C_k^{\star},~k=1,...,K.\\ 
\end{array}
\right.
\ee
\subsection{Bipartite Enhanced Configuration Model.}\label{subsec:bipecm}
The only difference with the Weighted model described in Appendix \ref{subsec:bipwcm} is 
the addition of the constraints on the number of degrees for each node. Before proceeding, 
we have thus to add some additional definitions. 

The binary projection of $X_{n,k}$ is defined as the matrix $Y_{n,k}={\mathds 1}_{X_{n,k}>0}$. 
Accordingly, the number  $D_n^{\textrm{row}}$
of assets in which the $n$-th bank invests and the number $D_k^{\textrm{col}}$ of banks that own the 
$k$-th asset class are computed as
\be\label{eq:degr}
D_n^{\textrm{row}}\tonde{\mathbf{X}} = \sum_{k=1}^{K}Y_{n,k},~~~~~~~D_k^{\textrm{col}}\tonde{\mathbf{X}} = \sum_{n=1}^{N}Y_{n,k},
\ee
where the capital letter $D$ stands for {\it degree},
as it is common practice in network theory\footnote{We have chosen 
not to use a different notation for bank and asset degrees since, although they have an immediate economic interpretation, 
they are not frequently used in economic analysis. In fact, while $A_n$ and $C_k$ are typically publicly available, the degree sequences
$D_n^{\textrm{row}}$ and $D_k^{\textrm{col}}$ are instead difficult to trace.}.

The maximization problem for the BIPECM case is hence 
stated as

\begin{equation}
\begin{aligned}
& \underset{\mathbb{P} }{\text{max}}
& & S\tonde{\mathbb{P}\quadre{\mathbf{X}}} \\
& \text{s.t.} & & \sum_{\mathbf{X}\in\mathcal{X}}\mathbb{P}\quadre{\mathbf{X}}=1\\
& & & \mathbb{E}_{\mathcal{X}}\quadre{A_n}=A_n^{\star},~n=1,...,N,\\
& & & \mathbb{E}_{\mathcal{X}}\quadre{D_n^{\textrm{row}}}=D_n^{\textrm{row}^{\star}},~n=1,...,N,\\
& & & \mathbb{E}_{\mathcal{X}}\quadre{C_k}=C_k^{\star},~k=1,...,K,\\
& &  &\mathbb{E}_{\mathcal{X}}\quadre{D_k^{\textrm{col}}}=D_k^{\textrm{col}^{\star}},~k=1,...,K.\nonumber
\end{aligned}
\end{equation}
 With an obvious extension of the number of Lagrange multipliers, and since $D_n^{\textrm{row}}\tonde{\mathbf{X}}=\sum_{k=1}^{K}{\mathds 1}_{X_{n,k}>0}$ (similarly, $D_k^{\textrm{col}}\tonde{\mathbf{X}}=\sum_{n=1}^{N}{\mathds 1}_{X_{n,k}>0}$), we obtain 
\ba
Z_{\boldsymbol{\vartheta}}  
                                             & = &\sum_{\mathbf{X}\in\mathcal{X}}\prod_{n=1}^{N}\prod_{k=1}^{K}e^{-\tonde{X_{n,k}\,\tonde{\lambda_n+\eta_k}+{\mathds 1}_{X_{n,k}>0}\,\tonde{\rho_n+\delta_k}}}\accapo
                                             & = & \sum_{X_{1,1}=0}^{\infty}\cdots\sum_{X_{1,K}=0}^{\infty}\cdots\sum_{X_{n,1}=0}^{\infty}\cdots\sum_{X_{n,K}=0}^{\infty}\prod_{n=1}^{N}\prod_{k=1}^K e^{-\tonde{X_{n,k}\,\tonde{\lambda_n+\eta_k}+{\mathds 1}_{X_{n,k}>0}\,\tonde{\rho_n+\delta_k}}}\accapo
                                             & = & \prod_{n=1}^{N}\prod_{k=1}^K\sum_{X_{n,k}=0}^{\infty} e^{-\tonde{X_{n,k}\,\tonde{\lambda_n+\eta_k}+{\mathds 1}_{X_{n,k}>0}\,\tonde{\rho_n+\delta_k}}}\accapo
                                             & = & \prod_{n=1}^{N}\prod_{k=1}^K\tonde{1+e^{-\tonde{\rho_n+\delta_k}}\sum_{X_{n,k}=1}^{\infty}e^{-\tonde{X_{n,k}\,\tonde{\lambda_n+\eta_k}}}} \accapo
                                             & = & \prod_{n=1}^{N}\prod_{k=1}^K\tonde{1+\frac{e^{-\tonde{\rho_n+\delta_k}}}{1-e^{-\tonde{\lambda_n+\eta_k}}}-e^{-\tonde{\rho_n+\delta_k}}}.\nonumber
\ea
Defining $\varphi_n=e^{-\lambda_n}$, $\xi_k=e^{-\eta_k}$, $\psi_n=e^{-\rho_n}$ and $\gamma_k=e^{-\delta_k}$ we get
\be\label{eq:Zbipecm}
Z_{\boldsymbol{\vartheta}} = \prod_{n=1}^{N}\prod_{k=1}^K\frac{1-\varphi_n\,\xi_k\,\tonde{1-\psi_n\,\gamma_k}}{1-\varphi_n\,\xi_k}.
\ee
Finally, we get that  the probability mass function for the BIPECM is
\be\label{eq:pdf_BIPECM}
\mathbb{P}_{\boldsymbol{\vartheta}}\tonde{\mathbf{X}} =  \prod_{n=1}^{N}\prod_{k=1}^K\frac{\tonde{1-\varphi_n\,\xi_k}\,\tonde{\varphi_n\,\xi_k}^{X_{n,k}}\,\tonde{\psi_n\,\gamma_k}^{{\mathds 1}_{X_{n,k}>0}} }{1-\varphi_n\,\xi_k\,\tonde{1-\psi_n\,\gamma_k}}.
\ee
The determination of the Lagrange multipliers follows a procedure identical 
to that described for BIPWCM in Appendix \ref{subsec:bipwcm}, that is the expected values of $A_n\tonde{\mathbf{X}}$, $C_{k}\tonde{\mathbf{X}}$, $D_n^{\textrm{row}}\tonde{\mathbf{X}}$ and $D_k^{\textrm{col}} \tonde{\mathbf{X}}$ are computed as the partial derivative of $-\log\tonde{Z_{\boldsymbol{\vartheta}}}$ with respect to the corresponding Lagrange multiplier and then equated to the observed value.
This procedure produce the non-linear system of equations

\be\label{eq:nnlinsys_nipecm}
\left\{
\begin{array}{rll}
\sum_{k=1}^K\frac{\phi_n\,\xi_k\,\psi_n\,\gamma_k}{\tonde{1-\phi_n\,\xi_k}\,\tonde{1-\phi_n\,\xi_k\,\tonde{1-\psi_n\,\gamma_k}}}&=& A_n^{\star},~n=1,...,n,\\
 & & \\
\sum_{k=1}^K\frac{\phi_n\,\xi_k\,\psi_n\,\gamma_k}{1-\phi_n\,\xi_k\,\tonde{1-\psi_n\,\gamma_k}} &=&  D_n^{\textrm{row}^{\star}},~n=1,...,n,\\
 & & \\
\sum_{n=1}^N\frac{\phi_n\,\xi_k\,\psi_n\,\gamma_k}{\tonde{1-\phi_n\,\xi_k}\,\tonde{1-\phi_n\,\xi_k\,\tonde{1-\psi_n\,\gamma_k}}}&=& C_k^{\star},~k=1,...,K,\\
 & & \\
\sum_{n=1}^N\frac{\phi_n\,\xi_k\,\psi_n\,\gamma_k}{1-\phi_n\,\xi_k\,\tonde{1-\psi_n\,\gamma_k}} &=&  D_k^{\textrm{col}^{\star}},~k=1,...,K.\\
 & & 
\end{array}
\right.
\ee

\section{Comparison of Reconstruction Methods under Different Shocks}\label{subsec:hetShockApp}
\begin{figure}[t]
\begin{center}

\includegraphics[width = 0.49\textwidth]{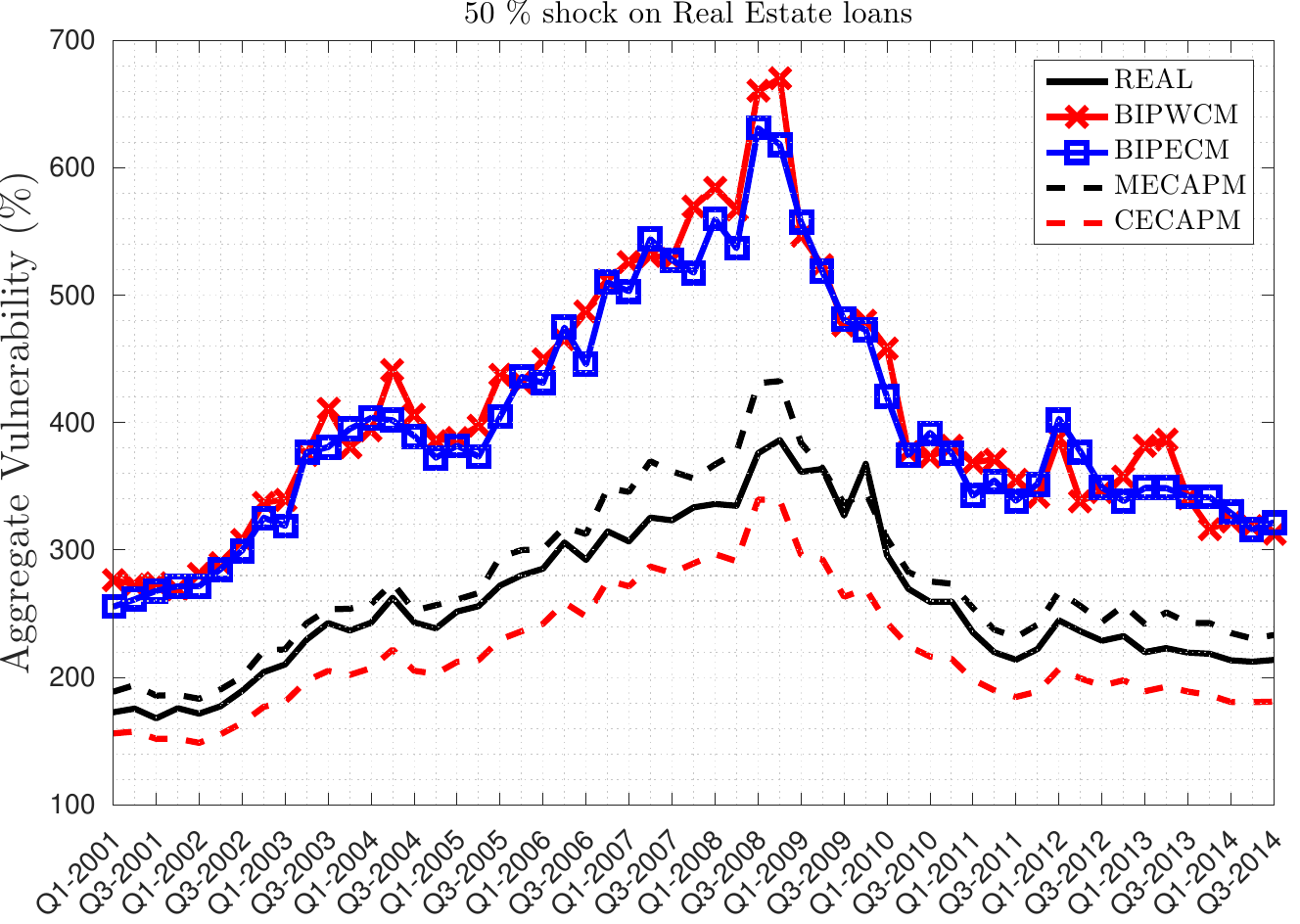} 
\includegraphics[width =  0.49\textwidth]{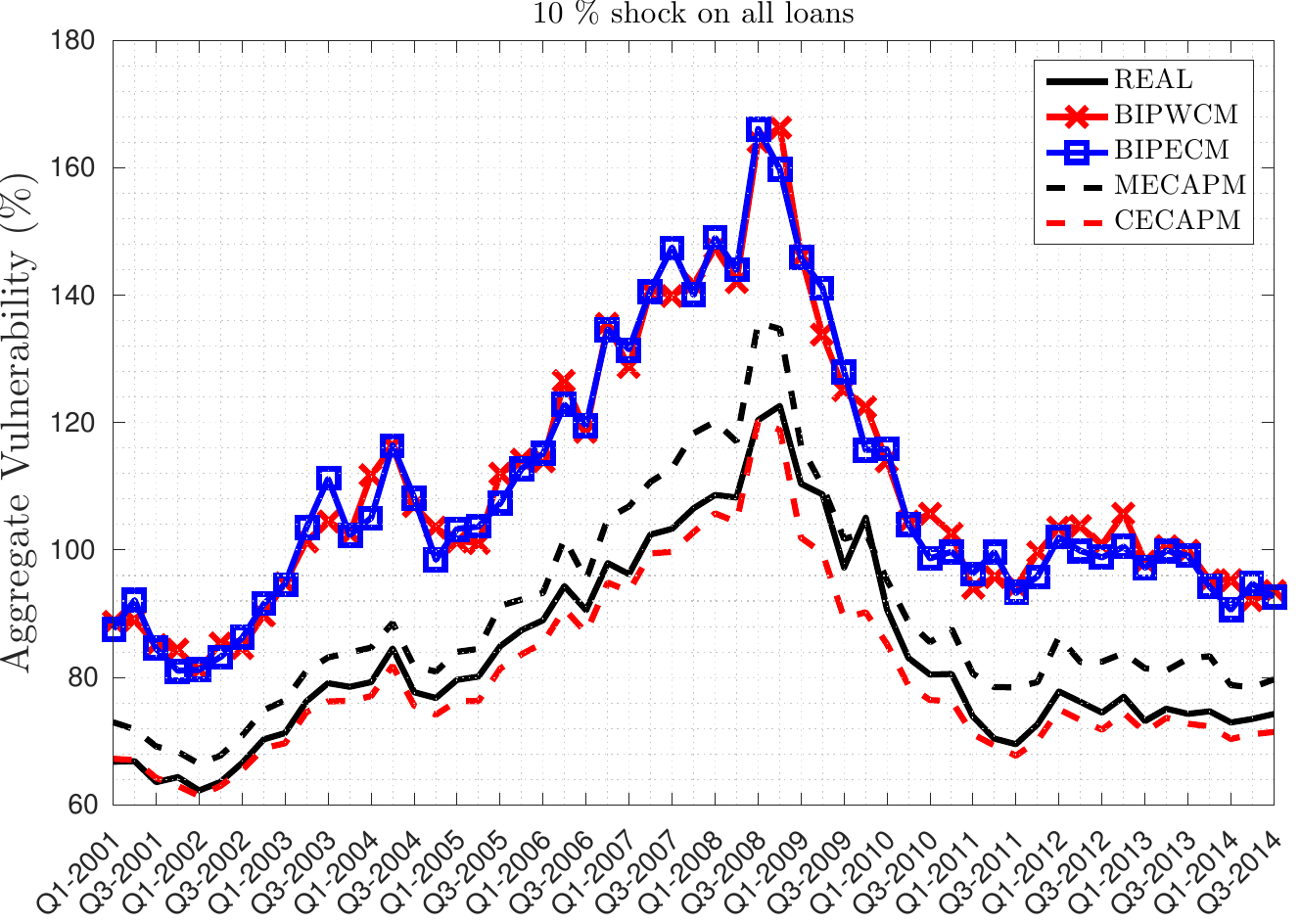} \\
\includegraphics[width = 0.49\textwidth]{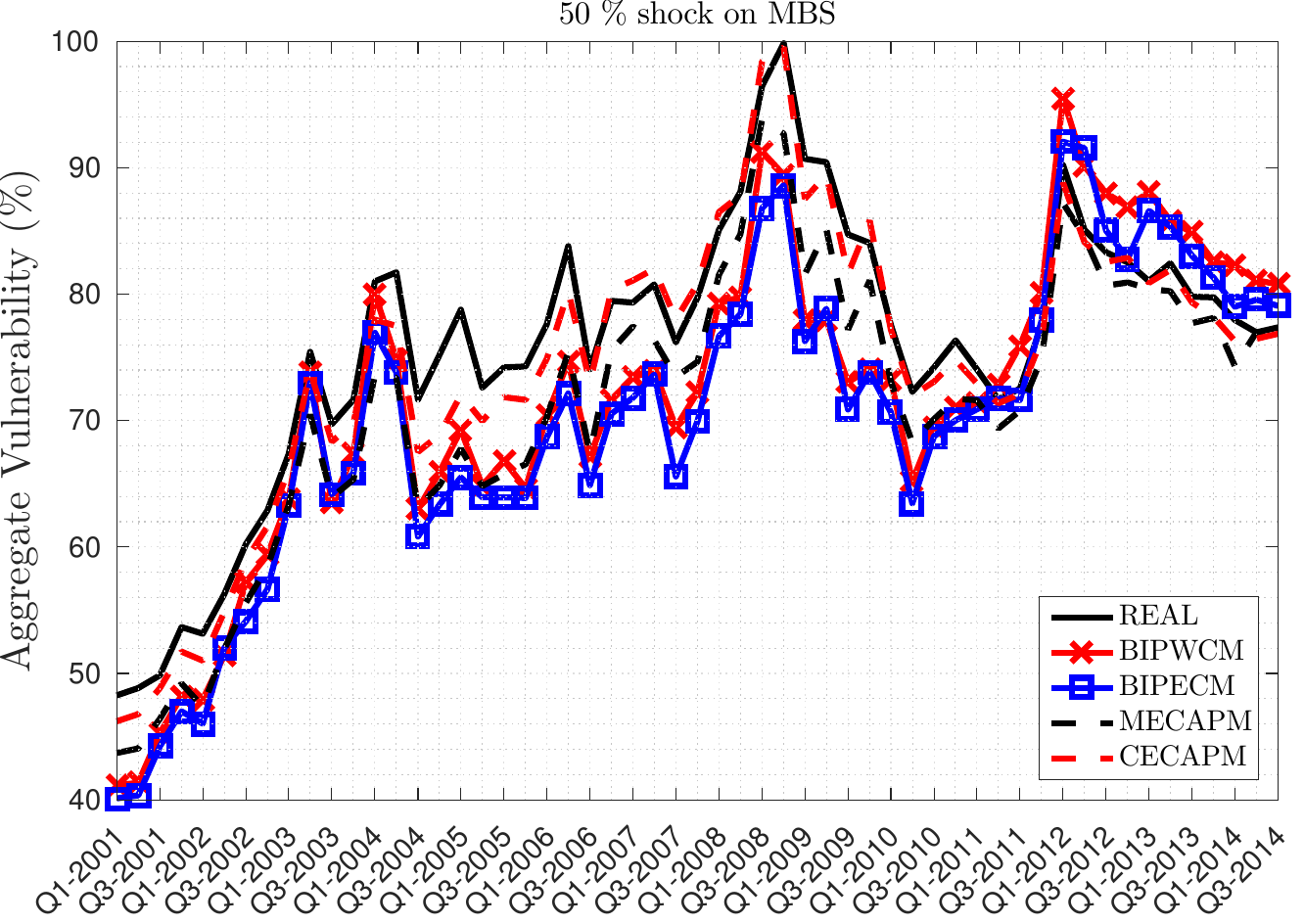} 
\includegraphics[width =0.49 \textwidth]{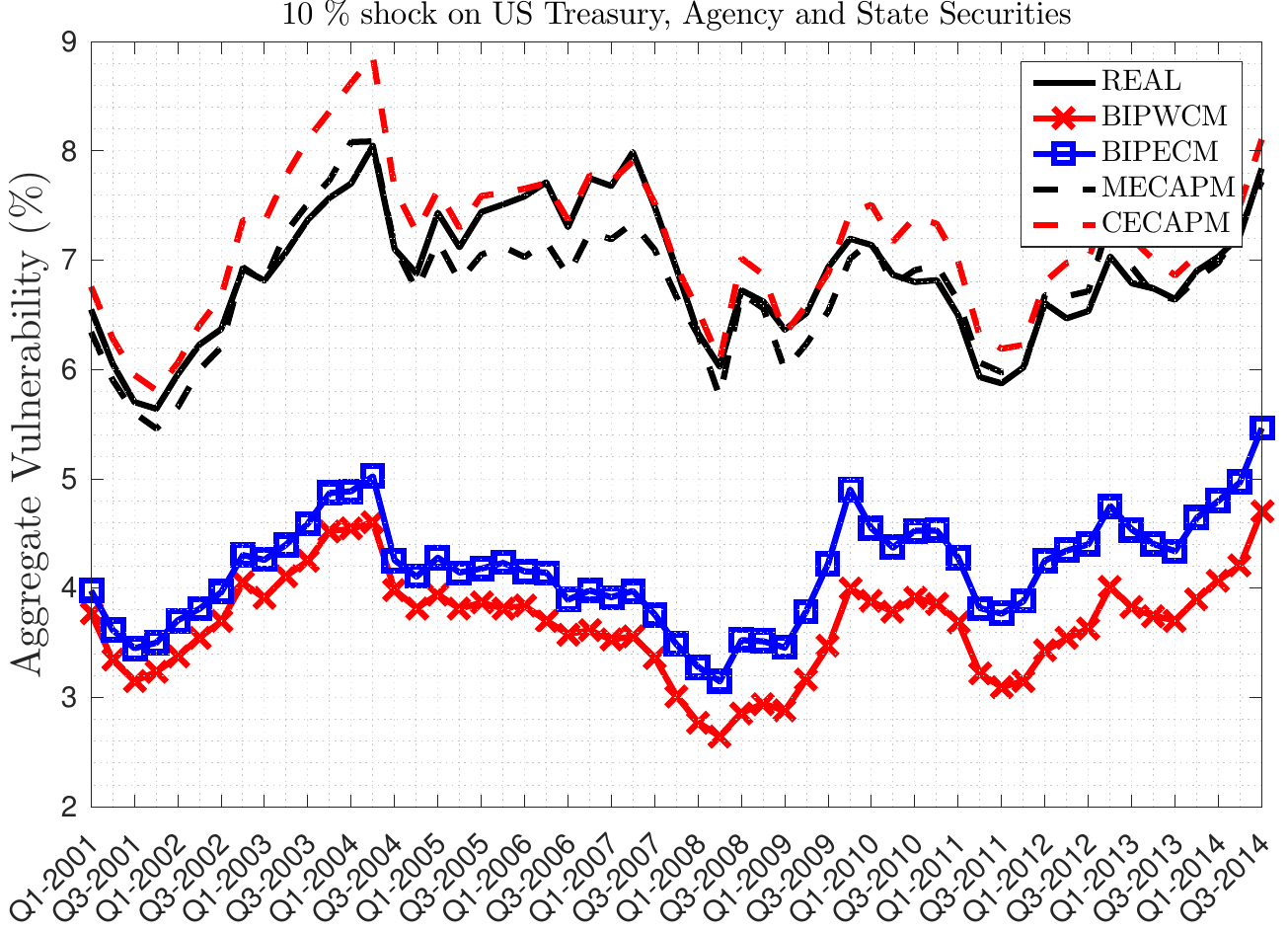} \\
 \caption{Aggregate vulnerability under different shock scenarios. Each panel reports the AV obtained from the full knowledge of portfolios composition and those obtained using the four reconstruction methods considered.}\label{fig:othershocksAllModels}
\end{center}
\end{figure}

In this appendix we report additional comparisons between cross entropy and maximum entropy ensemble methods for systemic risk assessment.  In Figure \ref{fig:othershocksAllModels} we show the performances of the four methods presented in the paper in reconstructing Aggregate Vulnerability, for four different shock scenarios. In all cases the CECAPM and MECAPM estimation outperform the other two and track quite closely the AV obtained from the full knowledge of portfolios composition. We therefore conclude that our result is not due to the uniform shock assumption, but is more generically applicable.

We then compare the performances of the different methods in assessing individual bank's quantities, it Indirect Vulnerability and Systemicness. In each quarter we have from $N=6,500$ to $N=9,000$ values of relative errors for each metric. To visualize the result we plot the median of the relative error and as a measure of dispersion we use the interquartile range, i.e. the difference between the upper and lower quartiles\footnote{We choose these metrics because they are more 
robust and less sensitive to outliers.}. Clearly a median well centered around zero is an indication that the estimator is unbiased. 
 
\begin{figure}[t]
\begin{center}
\includegraphics[width=0.75\textwidth]{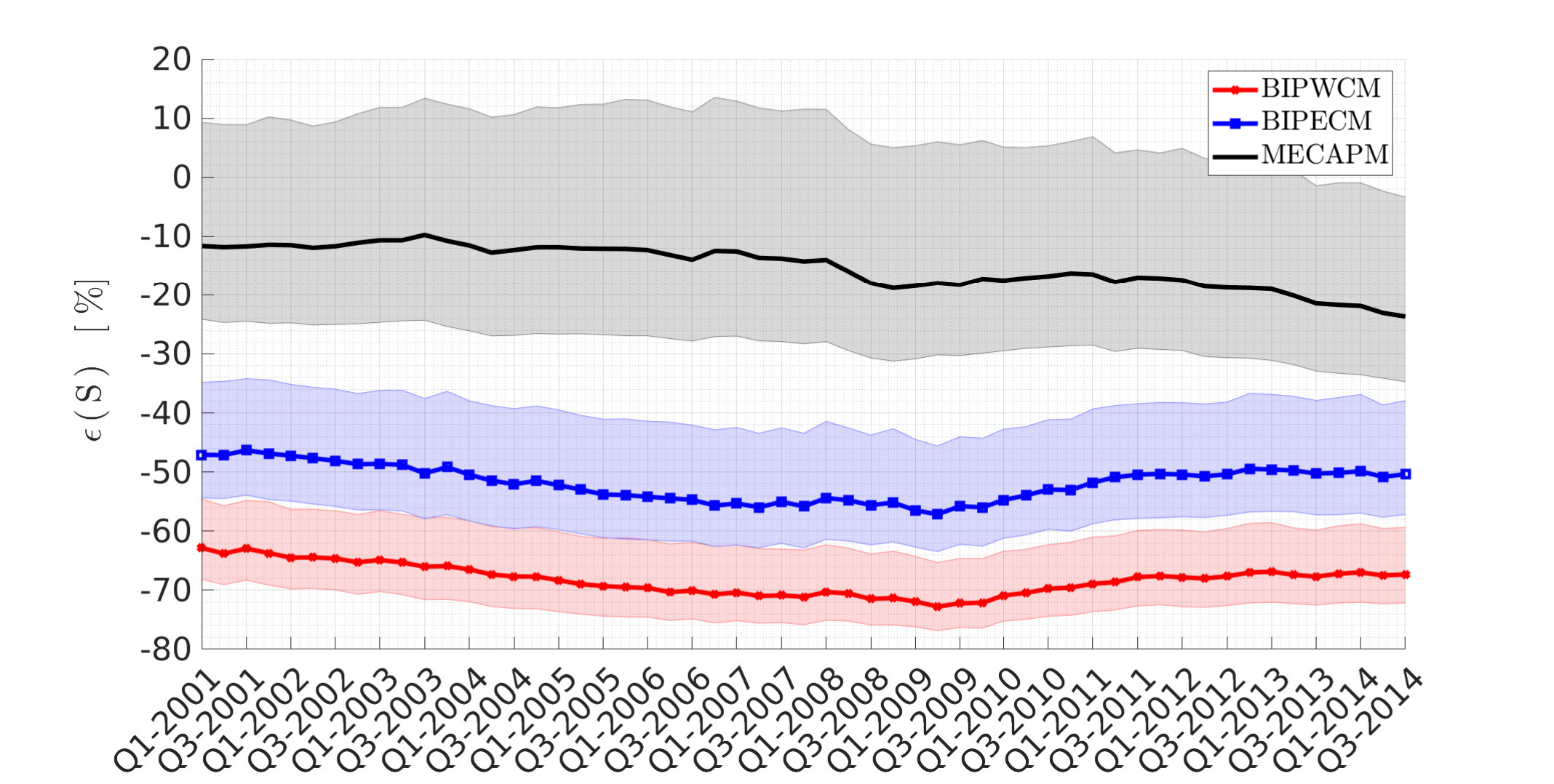}
\medskip
\includegraphics[width=0.75\textwidth]{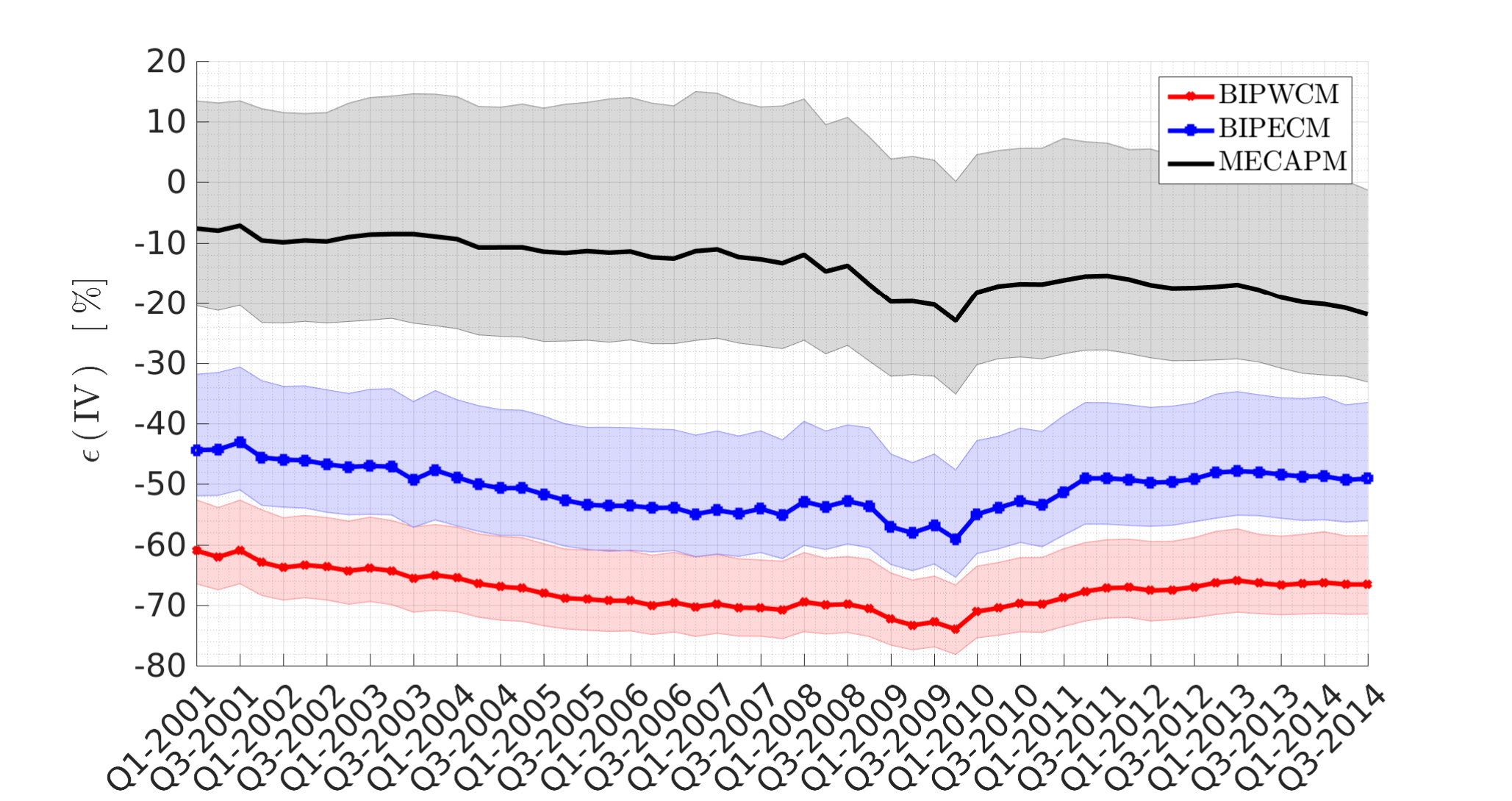}
\caption{\small Time series of the relative error (see Eq. \ref{eq:perc_error}) of bank systemicness (top panel) and indirect vulnerability (bottom panel) with respect to real data as estimated by the three ensembles BIPWCM (red and squares), BIPECM (blue and circles), and MECAPM (grey and dashed line). The thick lines indicate the median and the colored areas the interquartile range.}\label{fig:sysreconAllMethods}
\end{center}
\end{figure}

Figure \ref{fig:sysrecon} shows the results for bank systemicness (top panel) and indirect vulnerability (bottom panel). The three different colors refer to the three different ensembles (see the figure caption for more
details). We do not report the results for the CECAPM because it is indistinguishable from the MECAPM one and this fact can be understood by the argument in Eq. (\ref{eqapprox}).

We observe that for each quarter BIPWCM strongly underestimates individual bank systemicness and indirect vulnerability. The median relative error ranges roughly between $-60\%$ and $-70\%$ and the interquartile range is very far from zero. The estimator based on BIPECM (using the additional information on degrees) gives slightly better results, even if a strong underestimation is still present. The median relative error ranges roughly between  $-50\%$ and $-40\%$ and again the interquartile range is far from zero. On the contrary the estimator based on MECAPM (or CECAPM) performs much better. The median relative error never goes below $-20\%$ and almost always the interquartile range is centered around zero.\footnote{If instead we focus on the banks with higher systemicness or indirect vulnerability, the performances of the estimator based on MECAPM worsen. In particular,  for the quartile of banks with largest systemicness, the median percentage bias of the MECAPM estimator of systemicness is always between $-20\%$ and $-30\%$. Similarly, the median of the percentage bias in the estimation of indirect vulnerability via MECAPM is always between  $-20\%$ and $-35\%$. Nevertheless, the ranking among the three estimation methods remains unchanged.}
 
In summary, the estimates of systemicness and indirect vulnerably  for each 
single bank as provided by the CECAPM-implied matrix are almost identical to
those obtained as the corresponding expected values on the MECAPM ensemble. Besides, they 
are satisfactorily accurate and surely more reliable than those provided 
by standard maximum entropy ensembles. 

\section{Relation between CECAPM and the expected Values under MECAPM}\label{app:expform} 

This appendix is dedicated to derive explicit formulas for the expected values of bank
systemicness and indirect vulnerability under the MECAPM ensemble. Moreover we show 
that these expected values are well approximated by the corresponding values 
returned by the CECAPM-implied matrix.  
First note that, according to equation \eqref{eq:ens_geom}, the element $X_{n,k}$ is geometrically
distributed with parameter $p=\frac{1}{1+X^{\textrm{CAPM}}_{n,k}}$ and thus
$$
\E{X_{n,k}^2}-\E{X_{n,k}}^2 = \frac{1-p}{p^2} = X^{\textrm{CAPM}}_{n,k}\,\tonde{X^{\textrm{CAPM}}_{n,k}+1}.
$$
Hence the expected value of the systemicness of bank $n$ is given by\footnote{Remember that in our setting
all the illiquidity parameters are set to a common value $\ell$ except for the cash, which we assume to be the 
first class $k=1$, for which it is set to zero, whence the sum from $k=2$ to $k=K$.} 
\ba
\E{S_n\tonde{\mathbf{X}}} &=& \E{\Gamma_n}\,\frac{A_n^{\star}}{E^{\star}}\,B_n^{\star}\accapo
               &=& \ell\,\frac{B_n^{\star}}{E^{\star}}\,\sum_{k=2}^{K}\tonde{\E{X_{n,k}^2}+\sum_{m\neq n}\E{X_{n,k}}\,\E{X_{n,k}} }\accapo
               &=& \ell\,\frac{A_n^{\star}\,B_n^{\star}}{E^{\star}\,L^{\star}}\,\sum_{k=2}^KC_k^{\star}\,\tonde{X^{\textrm{CAPM}}_{n,k}+C_k^{\star}+1}.\nonumber
\ea
Note that since 
$$
S_n\tonde{\mathbf{X}^{\textrm{CAPM}}} = \ell\,\frac{A_n^{\star}\,B_n^{\star}}{E^{\star}\,L^{\star}}\,\sum_{k=2}^KC_k^{\star2},
$$
it is
$$
\E{S_n\tonde{\mathbf{X}}} = S_n\tonde{\mathbf{X}^{\textrm{CAPM}}} +  \ell\,\frac{A_n^{\star}\,B_n^{\star}}{E^{\star}\,L^{\star}}\,\sum_{k=2}^KC_k^{\star}\,\tonde{X^{\textrm{CAPM}}_{n,k}+1},
$$
whence
$$
\frac{\E{S_n\tonde{\mathbf{X}}}-S_n\tonde{\mathbf{X}^{\textrm{CAPM}}}}{S_n\tonde{\mathbf{X}^{\textrm{CAPM}}}}= \frac{\sum_{k=2}^KC_k^{\star}\,\tonde{X^{\textrm{CAPM}}_{n,k}+1}}{\sum_{k=2}^KC_k^{\star2}}.
$$
This expression can be rewritten as 
$$
\E{S_n\tonde{\mathbf{X}}}=S_n\tonde{\mathbf{X}^{\textrm{CAPM}}}\left(1+\frac{A_n^*}{L^*}+\frac{\sum_{k=2}^KC_k}{\sum_{k=2}^KC_k^{\star2}}\right),
$$
In order to evaluate the relative error above consider the simplified case in which all the
capitalization are almost equal $C_k^{\star}\approx C^{\star}=\frac{L^{\star}}{K}$, hence
\ba
\frac{\E{S_n\tonde{\mathbf{X}}}-S_n\tonde{\mathbf{X}^{\textrm{CAPM}}}}{S_n\tonde{\mathbf{X}^{\textrm{CAPM}}}}&\approx&\frac{\tonde{A_n^{\star}+K}}{L^{\star}}\accapo
& \approx&\frac{A_n^{\star}}{L^{\star}}\ll 1,\nonumber
\ea
where the last inequality follows from the fact that $L^{\star}=\sum_{n=1}^NA_n^{\star}$. 

Concerning the indirect vulnerability we have

\ba
\E{\textrm{IV}_n\tonde{\mathbf{X}}} & = & \ell\,\frac{\tonde{1+B_n^{\star}}}{A_n^{\star}}\,\E{\sum_{k=2}^KX_{n,k}\,\sum_{m=1}^NX_{m,k}\,B_m^{\star}} \accapo
 & = & \ell\,\frac{\tonde{1+B_n^{\star}}}{A_n^{\star}}\,\E{\sum_{k=2}^K\tonde{X_{n,k}^2\,B_n^{\star}+\sum_{m\neq n}X_{n,k}\,X_{m,k}\,B_m^{\star}}}\accapo
 & = & \ell\,\frac{\tonde{1+B_n^{\star}}}{A_n^{\star}}\,\sum_{k=2}^K\quadre{X^{\textrm{CAPM}}_{n,k}\,\tonde{2\,X^{\textrm{CAPM}}_{n,k}+1}B_n^{\star}+\sum_{m\neq n}X^{\textrm{CAPM}}_{n,k}\,X^{\textrm{CAPM}}_{m,k}\,B_m^{\star}}\accapo
 &=&\ell\,\frac{\tonde{1+B_n^{\star}}}{A_n^{\star}}\,\sum_{k=2}^K\quadre{\tonde{2\,X^{\textrm{CAPM}^2}_{n,k}+X^{\textrm{CAPM}}_{n,k}}B_n^{\star}+\sum_{m\neq n}X^{\textrm{CAPM}}_{n,k}\,X^{\textrm{CAPM}}_{m,k}\,B_m^{\star}}\accapo
 &=&\ell\,\frac{\tonde{1+B_n^{\star}}}{A_n^{\star}}\,\sum_{k=2}^K\quadre{\tonde{X^{\textrm{CAPM}^2}_{n,k}+X^{\textrm{CAPM}}_{n,k}}B_n^{\star}+\sum_{m=1}^NX^{\textrm{CAPM}}_{n,k}\,X^{\textrm{CAPM}}_{m,k}\,B_m^{\star}}\accapo
 & = &\ell\,\frac{\tonde{1+B_n^{\star}}}{A_n^{\star}}\,\sum_{k=2}^K\quadre{\tonde{\frac{A_n^{\star^2}\,C_k^{\star^2}}{L^{\star^2}}+\frac{A_n^\star\,C_k^\star}{L^\star}}B_n^{\star}+\frac{A_n^\star\,C_k^{\star^2}}{L^{\star^2}}\sum_{m=1}^NA_m^{\star}\,B_m^{\star}}.\nonumber
\ea
Now consider that $B_m^{\star}\approx \widebar{B}=10$, whence
\ba
\E{\textrm{IV}_n\tonde{\mathbf{X}}} &\approx& \ell\,\frac{\widebar{B}\,\tonde{1+\widebar{B}}}{A_n^{\star}}\,\sum_{k=2}^K\quadre{\frac{A_n^{\star^2}\,C_k^{\star^2}}{L^{\star^2}}+\frac{A_n^\star\,C_k^\star}{L^\star}+\frac{A_n^\star\,C_k^{\star^2}}{L^{\star}}}\accapo
&=&\ell\,\frac{\widebar{B}\,\tonde{1+\widebar{B}}}{L^{\star}}\,\sum_{k=2}^KC_k^\star\quadre{\frac{A_n^{\star}\,C_k^{\star}}{L^{\star}}+1+C_k^{\star}}.\nonumber
\ea
Now consider that
\ba
\textrm{IV}_n\tonde{\mathbf{X}^{\textrm{CAPM}}}& = & \ell\,\frac{\tonde{1+B_n^{\star}}}{A_n^{\star}}\,\sum_{k=2}^KX^\textrm{CAPM}_{n,k}\,\sum_{m=1}^NX^\textrm{CAPM}_{m,k}\,B_m^{\star}\accapo
&\approx & \ell\,\frac{\widebar{B}\,\tonde{1+\widebar{B}}}{L^{\star}}\,\sum_{k=2}^KC_k^{\star^2},\nonumber
\ea
whence
$$
\E{\textrm{IV}_n\tonde{\mathbf{X}}} \approx \textrm{IV}_n\tonde{\mathbf{X}^{\textrm{CAPM}}}+\ell\,\frac{\widebar{B}\,\tonde{1+\widebar{B}}}{L^{\star}}\,\sum_{k=2}^KC_k^\star\quadre{\frac{A_n^{\star}\,C_k^{\star}}{L^{\star}}+1}.
$$
Therefore
$$
\frac{\E{\textrm{IV}_n\tonde{\mathbf{X}}} - \textrm{IV}_n\tonde{\mathbf{X}^{\textrm{CAPM}}} }{\textrm{IV}_n\tonde{\mathbf{X}^{\textrm{CAPM}}} } = \frac{\sum_{k=2}^KC_k^\star\quadre{\frac{A_n^{\star}\,C_k^{\star}}{L^{\star}}+1}}{\sum_{k=2}^KC_k^{\star^2}}.
$$
Assume, again, for simplicity that $C_k\approx \frac{L}{K}$, thus
$$
\frac{\E{\textrm{IV}_n\tonde{\mathbf{X}}} - \textrm{IV}_n\tonde{\mathbf{X}^{\textrm{CAPM}}} }{\textrm{IV}_n\tonde{\mathbf{X}^{\textrm{CAPM}}} } \approx \frac{A_n^{\star}+K}{L^{\star}}\approx  \frac{A_n^{\star}}{L^{\star}}\ll 1
$$

 \clearpage
 {
\bibliographystyle{Chicago}
\bibliography{bib_survey}
}
\end{document}